\documentclass[preprint,12pt,sort&compress]{elsarticle}

\usepackage{amssymb}
\usepackage{amsmath}

\usepackage[colorlinks=true, allcolors=blue]{hyperref}
\usepackage{subfigure}
\journal{Annals of Physics}

\begin{document}

\begin{frontmatter}



\title{Wormhole Solutions and Pre-inflationary Epoch in $F(R, T)$ Gravity with Axion  Fields} 


\author[1]{Guo-He Li}
\ead{liguohe@stu.scu.edu.cn}

\author[1]{Yeqi Fang}
\ead{fangyeqi@stu.scu.edu.cn}
\author[2]{Yuchi Wu}

\author[1]{Jun Tao\corref{cor1}}
\ead{taojun@scu.edu.cn}

\cortext[cor1]{Corresponding author}
\affiliation[1]{organization={College of Physics, Sichuan University},
	addressline={}, 
	city={Chengdu},
	postcode={610065}, 
	state={},
	country={China}}

\affiliation[2]{organization={National Key Laboratory of Plasma Physics, Laser Fusion Research Center, CAEP},
	addressline={}, 
	city={Mianyang},
	postcode={621900}, 
	state={Sichuan},
	country={China}}
\begin{abstract}
This study investigates axion-dilaton wormhole solutions within the framework of $F(R,T)$ gravity to resolve the issue of insufficient inflationary e-folds in the no-boundary proposal. By examining both Giddings-Strominger and expanding wormhole solutions in asymptotically flat Euclidean spacetime, we demonstrate that the matter-geometry coupling induces complex dynamical oscillations in the scale factor and the dilaton field. These complex oscillatory modes significantly reduce the Euclidean action compared to standard general relativity, consequently enhancing the nucleation probability. Furthermore, we extend this theoretical setup to a ``wineglass'' half-wormhole model in Euclidean Anti-de Sitter (EAdS) spacetime. Our analysis yields a specific constraint on the coupling parameter. This constraint introduces an unstable maximum in the potential and simultaneously decreases the action. The presence of this unstable maximum fundamentally alters the probability distribution of initial states, rendering the evolution of universes from high-potential regions far more probable. Consequently, this approach significantly increases the likelihood of long-lasting inflation, providing a theoretical pathway to reconcile the no-boundary proposal with astronomical observations requiring sustained cosmic expansion.
\end{abstract}

\begin{graphicalabstract}
\end{graphicalabstract}

\begin{highlights}
	\item F(R,T) gravity induces complex oscillations in axion-dilaton wormholes.
	\item Enhanced oscillatory modes reduce Euclidean action to boost nucleation.
	\item Modified coupling yields an unstable potential maximum in EAdS spacetime.
	\item This mechanism favors prolonged inflation, solving no-boundary defects.
\end{highlights}


\end{frontmatter}



\section{INTRODUCTION}
For Einstein's theory of gravitation, the inflation emerged as the standard description for the early universe, successfully addressing critical cosmological problems including the flatness problem, the horizon problem~\cite{Guth:1980zm,Albrecht:1982wi,Linde:1981mu}, etc. It assumes that the universe underwent a period of super-exponential expansion in its very early stage, causing the initial perturbations to be rapidly amplified within a short period of time. This mechanism effectively accounts for the large-scale structure and anisotropy observed in the cosmic microwave background (CMB)~\cite{Guth:1982ec,Hawking:1982cz,Starobinsky:1982ee}.

The precise conditions that triggered cosmic inflation remain an unresolved question. Prior to inflation, the curvature and matter density of the universe approached the Planck scale, where quantum gravity effects become both significant and unavoidable, necessitating effective theories in quantum gravity such as the Wheeler-DeWitt (WDW) equation to investigate these conditions~\cite{DeWitt:1967yk,Halliwell:1988ik}. This equation yields multiple solutions, requiring the application of appropriate boundary conditions to select the physically relevant ones. Two prominent theories are the Hartle-Hawking no-boundary proposal~\cite{Hartle:1983ai,Hartle:2007gi}  and Vilenkin’s tunneling proposal~\cite{Vilenkin:1982de,Vilenkin:1983xq}. The no-boundary proposal asserts that the universe originated from a geometry without boundaries, predicting an inflationary perturbation spectrum that approximates a Gaussian distribution. However, its probability weight formula indicates that a smaller inflationary potential $V_{0}$ corresponds to a higher probability of universe creation. 
This implies that inflation with shorter duration and fewer e-folds is more probable, which conflicts with the observationally supported prolonged inflation~\cite{Maldacena:2024uhs,Lehners:2023yrj}. Research into Euclidean wormhole geometries offers a new perspective to address this challenge~\cite{Hawking:1987mz,Hawking:1988ae,Coleman:1988cy,Giddings:1988cx,Giddings:1988wv,DeFalco:2021ksd,Dai:2018vrw,Simonetti:2020ivl,Dai:2019mse,DeFalco:2020afv,Hong:2021qnk}. Driven by these cosmological implications, recent research has increasingly focused on constructing stable and traversable wormhole geometries within extended theoretical frameworks~\cite{javed2025wormholes,turimov2025exact,malik2026effect,dzhunushaliev2026wormholes,yousaf2026repulsive,Betzios:2026rbv,Lavrelashvili:2026zsw}.

Previous studies by Giddings and Strominger (GS) showed that a massless dilaton coupled to an axion field could support such structures; however, these solutions typically evolve into contracting ``baby universes'' upon analytic continuation to a Lorentzian spacetime, failing to describe our expanding universe~\cite{Giddings:1987cg}.
Recent investigations have revealed that a massive dilaton field broadens the range of possibilities, allowing for configurations that can produce expanding baby universes~\cite{Andriolo:2022rxc,Jonas:2023ipa}. A particularly promising class of solutions, known as ``wineglass'' wormholes, has emerged from the study of Euclidean axion-dilaton systems~\cite{Hebecker:2018ofv,Aguilar-Gutierrez:2023ril,Lavrelashvili:1988un,Bai:2023woh,Liu:2025foo,Betzios:2024oli,Betzios:2024iay}. These theories, whose Euclidean past is characterized by an asymptotically anti-de Sitter (AdS) boundary, can be analytically continued to describe an expanding universe. As recently proposed by Betzios and Papadoulaki, the wave function derived from the corresponding Euclidean path integral naturally evolves into an inflationary cosmology~\cite{Betzios:2024oli}. Axion wormholes thus provide a physically motivated framework for the initial conditions of inflation that may resolve the shortcomings of the no-boundary proposal.

Further advancements include the study of charged wormholes, where the inclusion of an electromagnetic field can enhance the probability weighting in certain regimes, potentially favoring prolonged expansion~\cite{Lan:2024gnv}. Nevertheless, these models have their own challenges. Within standard gravity, the Euclidean action of the traditional no-boundary state often remains lower than that of the charged wormhole, giving the former a probabilistic advantage. Furthermore, the viability of these Euclidean AdS (EAdS) wormhole models depends critically on a delicate balance between the axion charge, $Q$, and the inflationary potential, $V_0$. This fine-tuning problem, which represents a fundamental limitation for axion wormhole models in general relativity, strongly motivates the exploration of modified gravity theories that naturally introduce additional degrees of freedom.

To this end, we turn to modified gravity, where various extensions to General Relativity have been proposed, such as $F(R)$, $F(T)$, and $F(G)$ gravity~\cite{Odintsov:2019mlf,Oikonomou:2018npe,Nojiri:2017qvx,Malik:2022suu,Odintsov:2020nwm,Campa:2015bpr,Tayde:2022lxd,Zhong:2018tqn}. Among these, $F(R,T)$ gravity, which extends the gravitational action to depend on both the Ricci scalar $R$ and the trace of the stress-energy tensor $T$, is particularly compelling for our purposes~\cite{Harko:2011kv,Alvarenga:2013syu,Liang:2025hzr}. This theory offers the flexibility to address multiple cosmological puzzles simultaneously, from inflationary dynamics to dark energy~\cite{Sun:2015yga,Zaregonbadi:2016xna}. Although the theoretical self-consistency and validity of this framework on local scales remain debated and require verification through Solar System tests~\cite{Lacombe:2023pmx,Bertini:2023pmp}, recent studies demonstrate that it exhibits remarkable robustness and aligns closely with observational data on cosmological scales—the regime most relevant to the early universe~\cite{Harko:2024sea,Nagpal:2018mpv,Bose:2022xoc,Nashed:2023pxd,Jeakel:2023hss,Ashmita:2022swc,Deb:2022hna,Gamonal:2020itt}. This makes it a well-motivated framework for constructing stable wormhole geometries and investigating pre-inflationary physics~\cite{Moraes:2019pao,Elizalde:2018arz,Moraes:2017rrv,Elizalde:2018frj,Moraes:2017dbs,Sahoo:2017ual,Sahoo:2018kct,Azizi:2012yv,Sahoo:2019aqz}.

In this paper, we address the aforementioned limitations of axion-wormhole models by embedding them within the framework of $F(R,T)$ gravity. The additional degrees of freedom supplied by this theory offer a new mechanism to modulate the Euclidean action and throat geometry, potentially alleviating the fine-tuning problem and enhancing the probability of prolonged inflation. Specifically, this study systematically investigates two key categories of axion-dilaton wormhole solutions in $F(R,T)$ theory: the Giddings-Strominger (GS) type and the expanding type. The paper is organized as follows: Section~\ref{sec:2} develops the equations for the system under reflection symmetry boundary conditions. Section~\ref{sec:3} presents numerical solutions for both wormhole types and investigates the dependence of the throat geometry on the model's coupling parameters. In Section~\ref{sec:4}, we connect these solutions to inflationary cosmology by analyzing their implications for the probability of cosmic creation. Finally, Section~\ref{sec:5} concludes with a synthesis of our findings and a discussion of the physical constraints on the theory's parameters.

\section{$F\left(R,T\right)$ gravity coupled with axion}\label{sec:2}

In this study, we consider the Euclidean action for $F(R,T)$ gravity coupled to an axion and a dilaton/scalar $\phi$, which reads~\cite{Andriolo:2022rxc,Harko:2011kv}
\begin{equation}
	S_E=\int\mathrm{d}^{4}x\sqrt{g}\left(-\frac{1}{2\kappa}F(R,T)+\frac{1}{2}\nabla_{\mu}\phi\nabla^{\mu}\phi+V(\phi)\right.\left.+\frac{1}{12f^{2}}e^{{-\beta\phi\sqrt{\kappa}}}H_{\mu\nu\rho}H^{\mu\nu\rho}\right),
	\label{e1}%
\end{equation}
where $\kappa\equiv8\pi G$, $\beta$ is the dilatonic coupling constant, the dilaton potential is $V(\phi)$, and $H_\mathrm{\mu\nu\rho}$ is the 3-form field strength of an axion field with coupling $f$. For $\beta\neq0$, the field $\phi$ represents a dilaton, whereas $\beta=0$ is a simple scalar field. The axion field strength $H=$d$B$ is the exterior derivative of a 2-form, satisfying the Bianchi identity $\nabla_{[\mu}H_{\nu\rho\sigma]}=0$.
$F\left(R,T\right)$ is an arbitrary well-behaved function of the Ricci scalar $R=g^{\mu\nu}R_{\mu\nu}$, where $R_\mathrm{\mu\nu}$ is the Ricci tensor, and the trace of the stress-energy tensor $T=g^{\mu\nu}T_{\mu\nu}.$ The stress-energy tensor $T_\mathrm{\mu\nu}$ is defined in terms of the variation of the matter
\begin{equation}
	T_{\mu\nu}= -\frac{2}{\sqrt{g}}\frac{\delta\left(\sqrt{g}L_{\mathrm{m}}\right)}{\delta g^{\mu\nu}},
	\label{e2}
\end{equation}
it can be simplified to,
\begin{equation}
	\begin{aligned}
		T_{\mu\nu} = & \frac{1}{2} g_{\mu\nu}\nabla_\alpha\phi\nabla^\alpha\phi - \nabla_\mu\phi\nabla_\nu\phi + g_{\mu\nu}V(\phi) - \frac{1}{2f^2}\mathrm{e}^{-\beta\phi\sqrt{\kappa}}H_{\mu\rho\sigma}H_\nu^{\rho\sigma} \\
		& + \frac{1}{12f^2}\mathrm{e}^{-\beta\phi\sqrt{\kappa}}H_{\gamma\rho\sigma}H^{\gamma\rho\sigma}g_{\mu\nu}.
	\end{aligned}
\end{equation}
Taking the trace of the stress-energy tensor yields,
\begin{equation}
	T =  \nabla_\alpha\phi\nabla^\alpha\phi + 4V(\phi) - \frac{1}{6f^2} \mathrm{e}^{-\beta\phi\sqrt{\kappa}} H^2.
\end{equation}
By performing a variation and partial integration of Eq. $\left(  \ref{e1}\right)$ with respect to the metric tensor $ g_{\mu\nu} $, the modified field equations of the $F(R, T) $ gravity theory are derived,
\begin{equation}
	\begin{split}
		F_R\left(R,T\right)R_{\mu\nu} - \frac{1}{2}F\left(R,T\right)g_{\mu\nu} + \left(g_{\mu\nu}\square - \nabla_\mu\nabla_\nu\right)F_R\left(R,T\right) \\
		= \kappa T_{\mu\nu} - F_T\left(R,T\right)T_{\mu\nu} - F_T\left(R,T\right)\Theta_{\mu\nu}, \label{e4}
	\end{split}
\end{equation}
where we have defined the partial derivatives of \( F \) as \( F_R \equiv {\partial F}/{\partial R} \) and \( F_T \equiv {\partial F}/{\partial T} \), \( \nabla_\mu \) and \( \square \equiv \nabla^\sigma \nabla_\sigma \) are the covariant derivative and the D'Alembert operator. The auxiliary tensor \( \Theta_{\mu\nu} \) is  defined as~\cite{Rosa:2022osy},
\begin{equation}
	\begin{aligned}
		\Theta_{\mu\nu}\equiv g^{\alpha\beta}\frac{\delta T_{\alpha\beta}}{\delta g^{\mu\nu}}&= -2 T_{\mu\nu} + g_{\mu\nu} L_{m} -2 g^{\alpha\beta} \frac{\partial^2 L_m}{\partial g^{\mu\nu} \partial g^{\alpha\beta}},\\
	\end{aligned}
\end{equation}
substituting the relevant expressions yields,
\begin{equation}
	\Theta_{\mu\nu} = 2 \nabla_\mu \phi \nabla_\nu \phi - g_{\mu\nu} \left( \frac{1}{2} \nabla_\alpha \phi \nabla^\alpha \phi + V(\phi) \right.\left. + \frac{1}{12 f^2} \mathrm{e}^{-\beta \phi \sqrt{\kappa}} H_{\gamma\rho\sigma}H^{\gamma\rho\sigma} \right).
\end{equation}

In this study, we adopt the functional form $F(R,T)=R+\lambda T$. Unlike non-linear models such as $F(R,T) = R + \alpha R^2 + \lambda T$, which introduce fourth-order derivatives and extra scalar degrees of freedom~\cite{Sahoo:2017ual,Sahoo:2019aqz}, this minimal choice avoids such theoretical complexities while capturing the essential matter-geometry coupling effects. Ultimately, this mathematically elegant framework has been extensively studied in the literature and provides a robust foundation for our subsequent analysis~\cite{Zubair:2025jyw,Naseer:2024egv,Rastgoo:2024oao}.
\begin{equation}
	R_{\mu\nu}-\frac{1}{2}g_{\mu\nu}\left(R+\lambda T\right)=\kappa T_{\mu\nu}-\lambda\left(T_{\mu\nu}+\Theta_{\mu\nu}\right). \label{e6}
\end{equation}
Simplify the field equation, 
\begin{equation}
	\begin{aligned}
		R+\lambda T = & -\lambda\left(4V(\phi)+ \frac{1}{3 f^2} \mathrm{e}^{-\beta \phi \sqrt{\kappa}} H_{\mu\nu\rho}H^{\mu\nu\rho}\right) \\
		& + \kappa\left( \nabla_\alpha\phi\nabla^\alpha\phi + 4V(\phi) - \frac{1}{6f^2} \mathrm{e}^{-\beta\phi\sqrt{\kappa}} H_{\mu\nu\rho}H^{\mu\nu\rho}\right), \\
		\left( \frac{\lambda}{\kappa} - 1 \right) \nabla_\mu \nabla^\mu \phi = & \left( \frac{2\lambda}{\kappa} - 1 \right) \frac{\partial V}{\partial \phi} + \frac{\beta \sqrt{\kappa}}{12 f^2} \left( \frac{\lambda}{\kappa} + 1 \right) \mathrm{e}^{-\beta \phi \sqrt{\kappa}} H_{\mu\nu\rho} H^{\mu\nu\rho}, \\
		\partial_{\mu} \left( \sqrt{g} \mathrm{e}^{-\beta \phi \sqrt{\kappa}} H^{\mu \rho \sigma} \right) = & 0.
	\end{aligned}
	\label{e8}
\end{equation}
We will focus on the following spherically symmetric and homogeneous ansatz~\cite{Henneaux:1986ht}. The metric takes the form
\begin{equation}
	\mathrm{d}s^2=\mathrm{d}\tau^2+a(\tau)^2\left[\mathrm{d}\chi^2+\sin^2\chi\left(\mathrm{d}\theta^2+\sin^2\theta\,\mathrm{d}\phi^2\right)\right],
\end{equation}
For the three-form field strength $H$, we impose the conditions that all mixed time-space components vanish, $H_{0ij}=0$, while the purely spatial components are given by $H_{ijk}=q\varepsilon_{ijk}$, where $q$ is a constant parameter and $\varepsilon_{ijk}$ is the Levi-Civita symbol. Under these conditions, the action simplifies after integration by parts to
\begin{equation}
	\begin{aligned}
		S_E = & 2\pi^{2} \int \mathrm{d}\tau \left[ -\frac{3a\dot{a}^{2}}{\kappa} - \frac{3a}{\kappa} + \left( 1 - \frac{\lambda}{\kappa} \right) \frac{a^{3}\dot{\phi}^{2}}{2} + \left( 1 - \frac{2\lambda}{\kappa} \right) a^{3}V \right. \\
		& \left. + \left( 1 + \frac{\lambda}{\kappa} \right) \frac{N^{2}}{a^{3}} e^{-\beta\phi\sqrt{\kappa}} \right] + 2\pi^{2} \int \mathrm{d}\tau \frac{\mathrm{d}}{\mathrm{d}\tau} \left( \frac{3a^{2}\dot{a}}{\kappa} \right),
		\label{e11}
	\end{aligned}
\end{equation}
where $N^2\equiv\frac{q^2}{2f^2}$, varying the action yields the following equations of motion in the spherically symmetric ansatz:
\begin{equation}
	\begin{aligned}
		2a\ddot{a}+\dot{a}^2-1+\kappa a^2\left((1-\frac{\lambda}{\kappa}) \frac{\dot{\phi}^2}{2}+(1-\frac{2\lambda}{\kappa}) V(\phi)\right)-(1+\frac{\lambda}{\kappa}) \frac{\kappa N^2}{a^4}\mathrm{e}^{-\beta\phi\sqrt{\kappa}} &= 0, \\
		\dot{a}^2 - 1 - \frac{\kappa a^2}{3} \left( (1-\frac{\lambda}{\kappa})\frac{\dot{\phi}^2}{2} - (1-\frac{2\lambda}{\kappa})V(\phi)\right) + (1+\frac{\lambda}{\kappa}) \frac{\kappa N^2}{3a^4} e^{-\beta \phi \sqrt{\kappa}} &= 0, \\
		\ddot{\phi} + 3 \frac{\dot{a}}{a} \dot{\phi} - \frac{1 - \frac{2\lambda}{\kappa}}{1 - \frac{\lambda}{\kappa}} \frac{\partial V}{\partial \phi} + \frac{1 +\frac{\lambda}{\kappa}}{1 - \frac{\lambda}{\kappa}} \frac{\beta \sqrt{\kappa} N^2}{a^6} e^{-\beta \phi \sqrt{\kappa}} &= 0.
	\end{aligned}
	\label{e13}
\end{equation}
Using the first equation in Eq. $\left(  \ref{e8}\right)$  , the on-shell action can be expressed as
\begin{equation}
	S_E=2\pi^2\int\mathrm{d}\tau \left[2(1+\frac{\lambda}{\kappa})\frac{2N^2\mathrm{e}^{-\beta\phi\sqrt{\kappa}}}{a^3}-(1-\frac{2\lambda}{\kappa})a^3V(\phi)\right],\label{e16}
\end{equation}
which is equivalent to the action Eq. $\left(  \ref{e11}\right)$  by using the equations of motion Eq. $\left(  \ref{e13}\right)$.

The Euclidean wormholes can be interpreted as tunneling events leading to the creation of  baby universes~\cite{Coleman:1988cy}. A regular wormhole at $ \tau = 0 $ is characterized by a finite spatial size $ a(0) = a_0 \neq 0 $ and a vanishing initial derivative of the scale factor, \( \dot{a}(0) = 0 \). For infinitesimal \( \tau \), it can be expanded as~\cite{Lavrelashvili:2026zsw,Jonas:2023ipa}:
\begin{equation}
	a(\tau) = a_0 + \frac{1}{2} \ddot{a}(0) \tau^2 + \mathcal{O}(\tau^4).
\end{equation}
By performing analytic continuation to Minkowski time through \( t = -i\tau \), the expression transforms into:
\begin{equation}
	a(t) = a_0 - \frac{1}{2} \ddot{a}(0) t^2 + \mathcal{O}(t^4).
\end{equation} 
For GS wormholes, the ``throat'' represents a minimum, implying $\ddot{a}(0) > 0$, which corresponds to a contracting universe. In contrast, to facilitate an expanding wormhole, it is required that $\ddot{a}(0) < 0$, indicating that the ``throat'' acts as a local maximum of the size function.

At $\tau = 0$, the second equation of  Eq. $\left(  \ref{e13}\right)$ reduces to the Friedmann constraint, which establishes a connection between the initial values of the scale factor $a_0$ and the scalar field $\phi_0$.
\begin{equation}
	\frac{3}{\kappa a_0^2} = \frac{\kappa-2\lambda}{\kappa}V(\phi_0) +\frac{\kappa+\lambda}{\kappa}\frac{N^2 e^{-\beta \sqrt{\kappa}\phi_0}}{a_0^6} 
\end{equation}
Simplify the equation by defining \( Q^2 = N^2 \mathrm{e}^{-\beta \sqrt{\kappa} \phi_0} \) and \( x = a_0^2 \),
\begin{equation}
	\frac{\kappa-2\lambda}{3}V(\phi_0)x^3-x^2+\frac{\kappa+\lambda }{3}Q^2=0. \label{e19}
\end{equation}
Therefore, the discriminant of the cubic equation given in Eq. (\ref{e19}) can be calculated as follows:
\begin{equation}
	\Delta=\frac{(\kappa+\lambda) Q^2}{3}\left[4-(\kappa+\lambda)(\kappa-2\lambda)^2 Q^2V^2(\phi_0) \right].
\end{equation}
When the discriminant \(\Delta > 0\), the equation \(x = a_0^2\) admits three real solutions, typically yielding two distinct positive roots for \(a_0\). These two positive roots correspond to the initial conditions for two physically different wormhole types: the contracting Giddings-Strominger type and the expanding type. The case \(\Delta < 0\) yields only one real root, so we do not consider it in our work. 
For $\Delta>0$, we can define an angle $\theta\in (0,\pi]$,
\begin{equation}
	\cos\theta=1-\frac{1}{2}(\kappa-2\lambda)^2(\kappa+\lambda)Q^2V^2(\phi_0),
\end{equation}
Then the three real roots of the cubic equation can be expressed as follows:
\begin{equation}
	x_i=\frac{1}{(\kappa-2\lambda) V(\phi_0)}\left(1+2\cos\frac{\theta-2\pi\cdot i}{3}\right) ~\mathrm{for~}i=0,1,2.
	\label{e22}
\end{equation}
The solution corresponding to \( i = 2 \) yields a negative value of \( x \) and is therefore discarded. We thus obtain four real solutions for \( a \), consisting of two positive and two negative roots. Since only positive values of \( a \) are physically meaningful, we retain exclusively the positive solutions for further analysis. Among these positive solutions, the larger one, denoted as \( a_{\text{max}} \), is bounded by
\begin{equation}
	\sqrt{\frac{2}{(\kappa-2\lambda) V(\phi_0)}} < a_{max} \leq \sqrt{\frac{3}{(\kappa-2\lambda) V(\phi_0)}}
	\label{e23}
\end{equation}
To ensure the equation corresponds to a physically sensible function, we impose the additional constraint \(\lambda < \kappa/2\). When \( Q \) $\to$ 0 or \( \lambda \) $\to$ \(-\kappa\), equality holds, and it approaches the size of the Hubble radius.

\section{Axion-dilaton wormholes in the flat Euclidean spacetime }\label{sec:3}
After obtaining the equations of motion and understanding the formation mechanisms of different baby universes, this section investigates two types of solutions within the framework of axion-dilaton modified gravity: GS wormhole and expanding wormhole. We will employ the potential  for the massive case,
\begin{equation}
	V(\phi)=\frac{1}{2}m^2\phi^2,
\end{equation}
where $m$ is the dilaton mass. As already demonstrated in recent literature~\cite{Jonas:2023ipa}, the adoption of the massive potential enables wormhole solutions to exist beyond classical critical coupling limits. It also introduces new geometric branches featuring multiple minima and oscillatory dynamics, thereby providing a significantly broader solution space for our investigation. And the Eq. $\left(  \ref{e13}\right)$ then becomes
\begin{equation}
	\begin{aligned}
		2a\ddot{a}+\dot{a}^2-1+\kappa a^2\left((1-\frac{\lambda}{\kappa}) \frac{\dot{\phi}^2}{2}+(1-\frac{2\lambda}{\kappa}) \frac{m^2\phi^2}{2}\right)-(1+\frac{\lambda}{\kappa}) \frac{\kappa N^2}{a^4}\mathrm{e}^{-\beta\phi\sqrt{\kappa}} &= 0 , \\
		\dot{a}^2 - 1 - \frac{\kappa a^2}{3} \left( (1-\frac{\lambda}{\kappa})\frac{\dot{\phi}^2}{2} - (1-\frac{2\lambda}{\kappa})\frac{m^2\phi^2}{2} \right) + (1+\frac{\lambda}{\kappa}) \frac{\kappa N^2}{3a^4} e^{-\beta \phi \sqrt{\kappa}} &= 0, \\
		\ddot{\phi} + 3 \frac{\dot{a}}{a} \dot{\phi} - \frac{1 - \frac{2\lambda}{\kappa}}{1 - \frac{\lambda}{\kappa}} m^2 \phi + \frac{1 + \frac{\lambda}{\kappa}}{1 - \frac{\lambda}{\kappa}} \frac{\beta \sqrt{\kappa} N^2}{a^6} e^{-\beta \phi \sqrt{\kappa}} &= 0 .
	\end{aligned}
	\label{e25}
\end{equation}
From the boundary terms in reduced action Eq. $\left(  \ref{e11}\right)$, we can get the initial conditions $\dot{a}(0) = 0 $ and $\dot{\phi}(0) = 0 $ at the wormhole neck \( \tau = 0 \)~\cite{Jonas:2023ipa}. Furthermore, the conditions in the asymptotic future (\(\tau_f \to \infty\)) are \( \dot{a}(\infty) = 1 \) and \( \phi(\infty) = 0 \), which imply an asymptotically flat Euclidean spacetime. 
\begin{equation}
	\quad \dot{a} (0)= 0, \quad \dot{\phi}(0) = 0,\quad \dot{a}(\tau \rightarrow \infty )=1, \quad \phi (\tau \rightarrow \infty )=0 .
	\label{e26}
\end{equation}
Next, we will utilize these conditions to find the wormhole solutions. The initial scale factor $a_0$ is determined by $\phi_0$ through Eq.~$\left(  \ref{e22}\right)$. For all solutions, the smaller positive value of \( a_0 \) means a local minimum of the scale factor. This indicates that these wormholes (GS) would lead to contracting universes. Conversely, the larger positive value of \( a_0 \) corresponds to a local maximum of the scale factor, indicating that such baby universes would undergo continuous expansion in Lorentzian time.

We can employ the shooting method to identify the appropriate initial value of the dilaton field, \(\phi_0\), that satisfies the boundary condition \(\phi(\tau \to \infty) = 0\). This method uses the asymptotic behavior of the dilaton field, where small perturbations in \(\phi_0\) cause \(\phi(\tau)\) to transition between diverging to positive infinity and negative infinity. By the intermediate value theorem, we infer the existence of a critical \(\phi_0\) at which \(\phi(\tau \to \infty) = 0\), thus yielding the desired solution. We selected the first and third equations from Eq. $\left(  \ref{e25}\right)$  and incorporated the boundary conditions from Eq. $\left(  \ref{e26}\right)$  to obtain $a_0$ and $\phi_0$, which allows us to explore the dynamics and stability of wormholes.

We first focus on the GS solutions that lead to the formation of contracting baby universes. This type of solution has been  addressed in recent literature. Previous studies have explored different parameter spaces for these wormhole solutions: Andriolo et al. examined the effects of varying $m$ while holding the parameter $N \equiv q/(\sqrt{2}f) = 1$ constant~\cite{Andriolo:2022rxc}, whereas Jonas et al. investigated the consequences of varying $N$ while fixing $m = 10^{-2}$~\cite{Jonas:2023ipa}. Our work distinguishes itself from these earlier studies by shifting the focus to the coupling parameter $\lambda$. To achieve this, we have set $\beta=1.2$, $m=0.01$, and $N=30000$,  allowing us to study the isolated impact of $\lambda$ on these wormhole solutions.

\begin{figure}[htbp]
	\centering
	\subfigure{\includegraphics[width=0.47\textwidth]{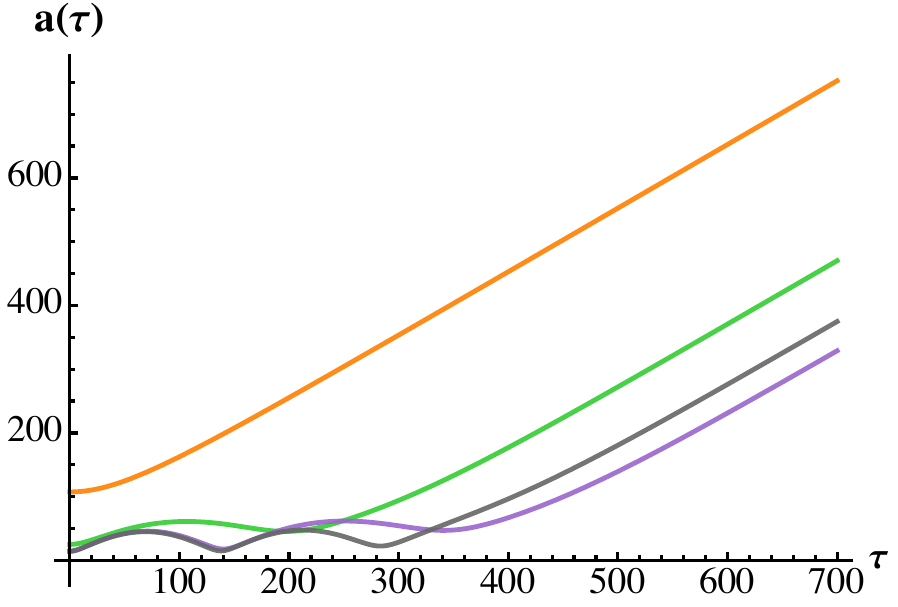}}
	\subfigure{\includegraphics[width=0.47\textwidth]{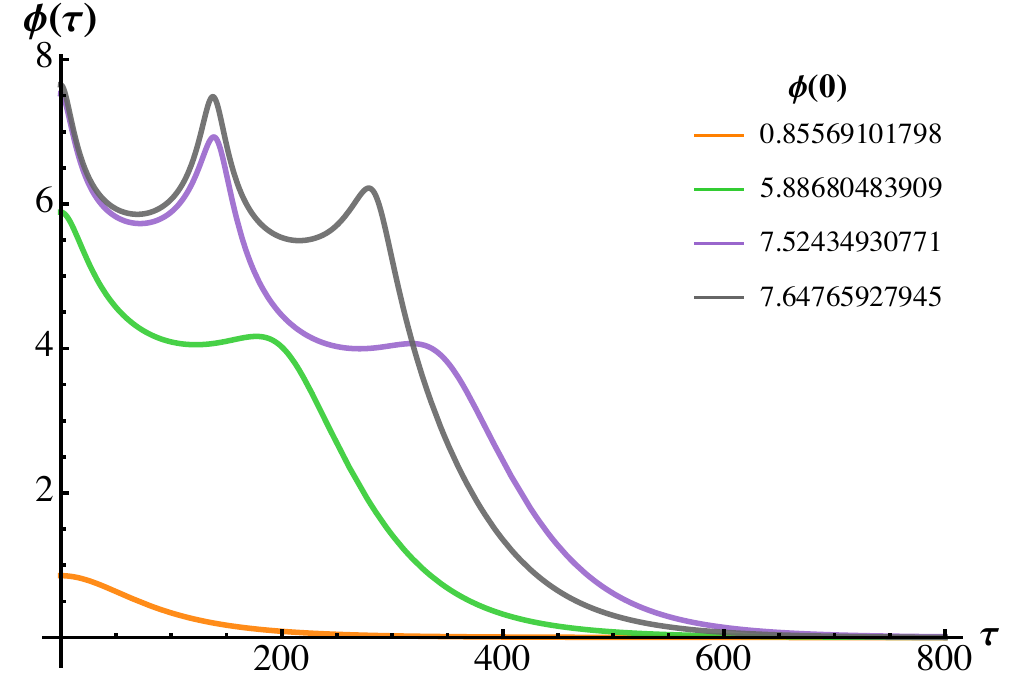}}
	\vspace{-0.4cm}
	\caption{Evolution of the GS wormhole solutions: the scale factor is shown on the left and the dilaton evolution on the right. All solutions are characterized by the same parameters: \(\kappa = 1\), \(\beta = 1.2\), \(N = 30000\), \(m = 0.01\), and \(\lambda = 0.1\). The individual solutions are distinguished by the initial values of the dilaton field, which are \(0.85569101798\),  \(5.88680483909\), \(7.52434930771\),and \(7.64765927945\), respectively. }
	\label{fig:lam0.1}
\end{figure}

The characteristics of the wormhole solution are depicted in Fig.~\ref{fig:lam0.1} with the parameter \(\lambda = 0.1\).  It is observed that as \(\phi_0\) increases, the evolution of the dilaton field becomes more complex, with distinct oscillatory behavior emerging. Specifically, both the scale factor and the dilaton field exhibit oscillations~\cite{Lee:2009bp}. For solutions with larger \(\phi_0\), the dilaton field \(\phi\) and the scale factor \(a\) display two maxima and two minima, indicating that the wormhole throat oscillates twice. This behavior is markedly different from the oscillatory bounce mechanism reported in work~\cite{Hackworth:2004xb}. The oscillations of the dilaton and the scale factor are more frequent and have larger amplitudes compared with $\lambda$=0~\cite{Jonas:2023ipa}.

\begin{figure}[htbp]
	\centering
	\subfigure{\includegraphics[width=0.49\textwidth]{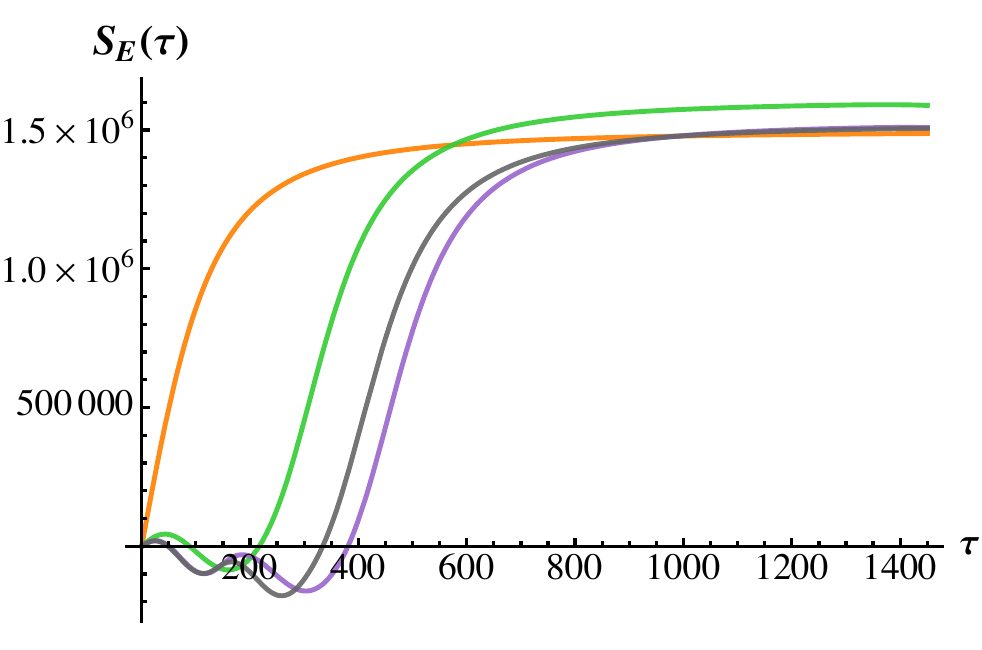}}
	\subfigure{\includegraphics[width=0.49\textwidth]{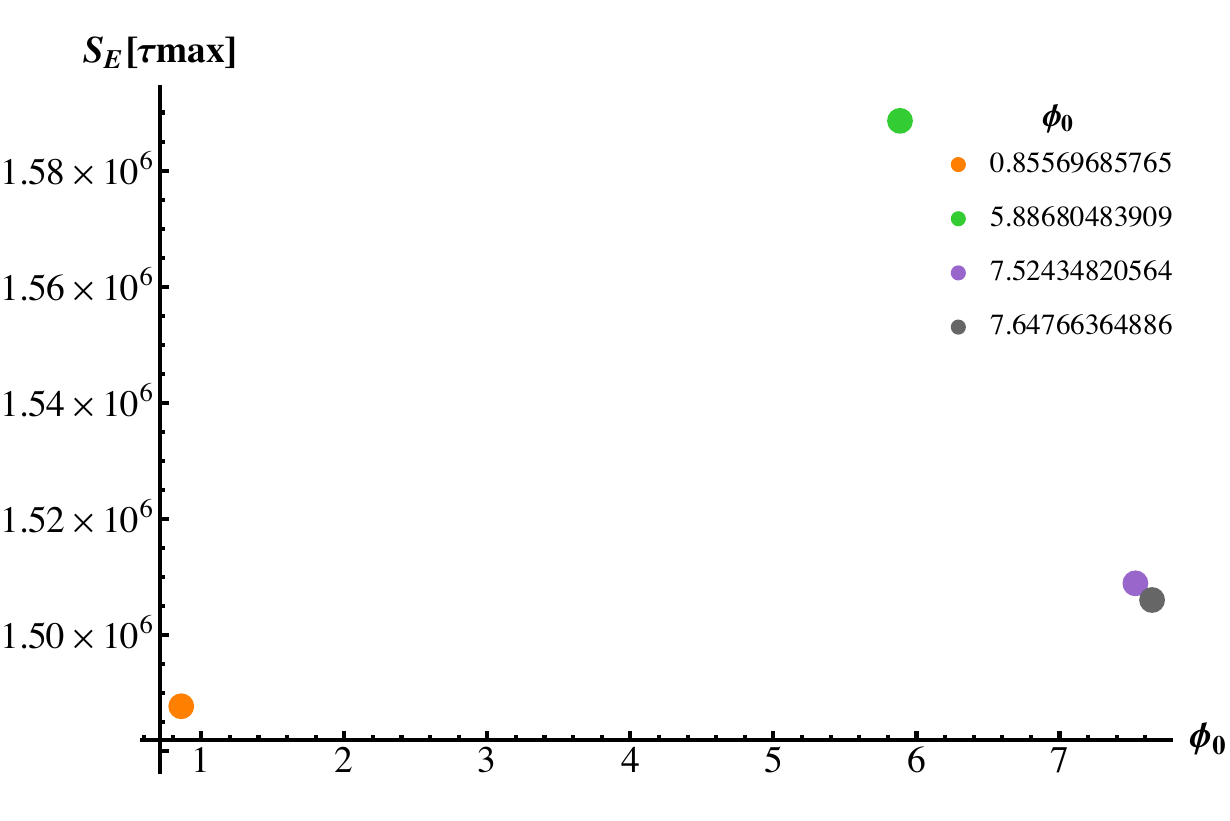}}
	\vspace{-0.4cm}
	\caption{The left plot shows the Euclidean action as a function of $\tau$, while the right plot displays the corresponding asymptotic values of the solutions in Fig.~\ref{fig:lam0.1}. Notably, the action does not exhibit a monotonic behavior with respect to \( \phi_0 \); instead, it begins to decrease with the introduction of additional oscillations. }
	\label{fig:lam0.1SE}
\end{figure}

Based on the scalar field and the dilaton field, we can calculate the corresponding Euclidean action using Eq. $\left(  \ref{e16}\right)$ as shown in Fig.~\ref{fig:lam0.1SE}. In particular, the second solution (the green line) exhibits an inflection point in the scale factor evolution, corresponding to a relatively larger action. The Euclidean action of these solutions tends to stabilize as time increases. This can be attributed to the evolution of the scale factor \(a\) and the dilaton field \(\phi\) towards flat space conditions at larger time scales \(\tau\). A key feature of these solutions is that the Euclidean action remains positive, which is consistent with the interpretation of these solutions as mediating the nucleation process of tunneling events in the baby universe. As the initial dilaton field $\phi_0$ increases, the evolution of the action becomes more complex, and may exhibit a transition from negative to positive values. For solutions with oscillatory behavior, especially those with two additional minima, the introduction of the coupling parameter leads to a reduction in the action compared to the situation $\lambda$=0~\cite{Jonas:2023ipa}. Assuming the nucleation probability per unit four-volume is approximately given by $e^{-2S_E/\hbar}$, it can be inferred that solutions with additional oscillations are more likely to occur. As indicated in the right panel of Fig.~\ref{fig:lam0.1SE}, the final action does not have a monotonic relationship with \(\phi_0\). Notably, solutions featuring a higher number of oscillations can yield a final action lower than that of the simplest, non-oscillatory solution. This suggests that complex oscillatory tunneling events can be probabilistically competitive, or even favored, over simpler ones. 

\begin{figure}[htbp]
	\centering
	\subfigure{\includegraphics[width=0.47\textwidth]{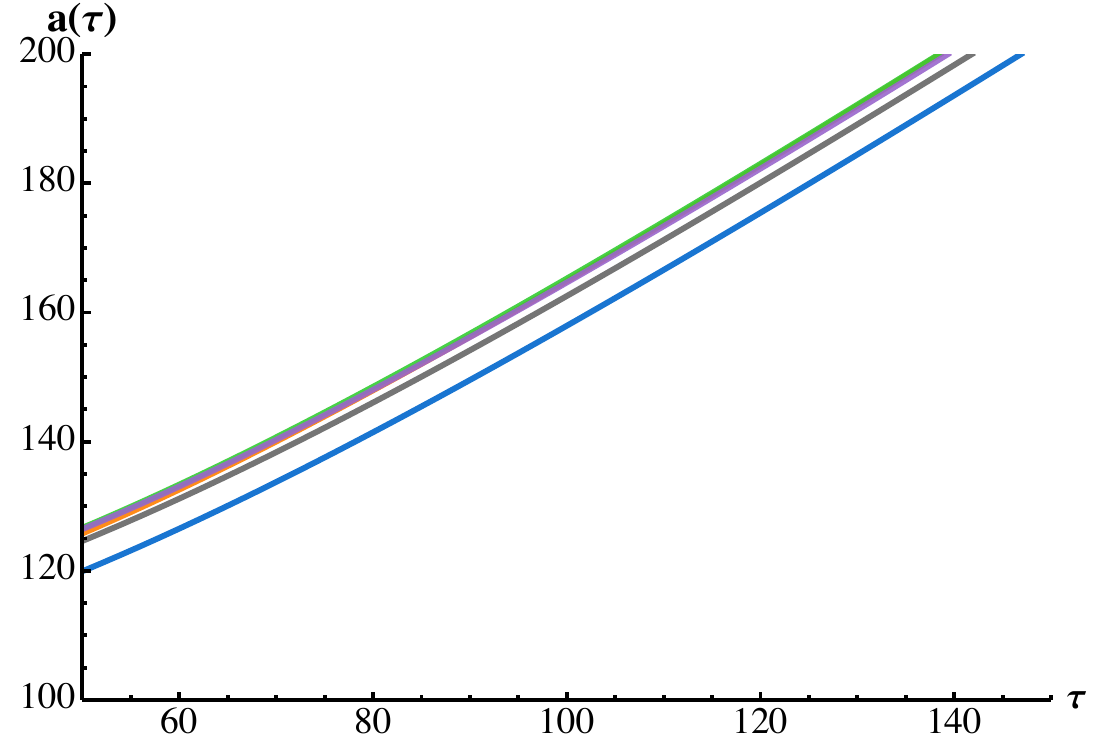}}
	\subfigure{\includegraphics[width=0.47\textwidth]{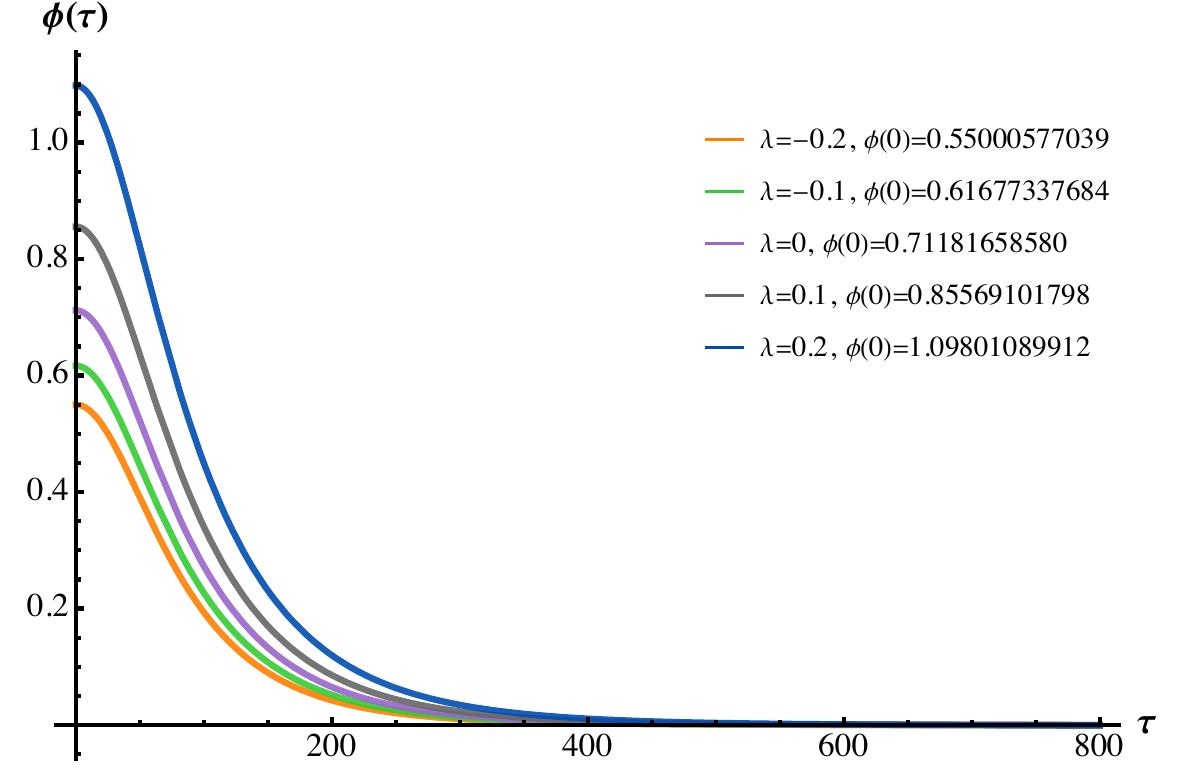}}
	\vspace{-0.4cm}
	\caption{Evolution of the wormhole under different values of \(\lambda\), with the scale factor and the dilaton field shown on the left and right, respectively. All solutions share the same parameters: \(\kappa = 1\), \(\beta = 1.2\), \(N = 30000\), and \(m = 0.01\). However, they exhibit different values of \(\lambda\), ranging from \(-0.2\) to \(0.2\) in increments of \(0.1\). The corresponding values of \(\phi_{0,\text{min}}\) are 0.55000577039, 0.61677337684, 0.71181658580, 0.85569101798, 1.09801089912. It is evident that \(\phi_{0,\text{min}}\) is positively correlated with \(\lambda\). }
	\label{fig:diflammin}
\end{figure}
\begin{figure}[htbp]
	\centering
	\subfigure{\includegraphics[width=0.49\textwidth]{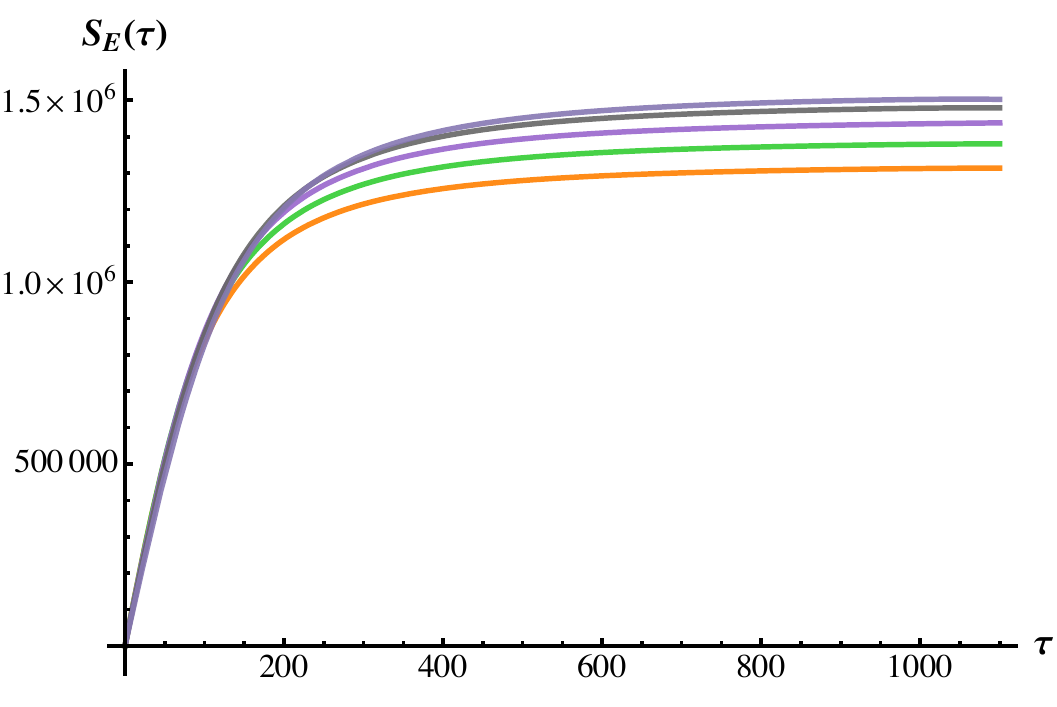}}
	\subfigure{\includegraphics[width=0.49\textwidth]{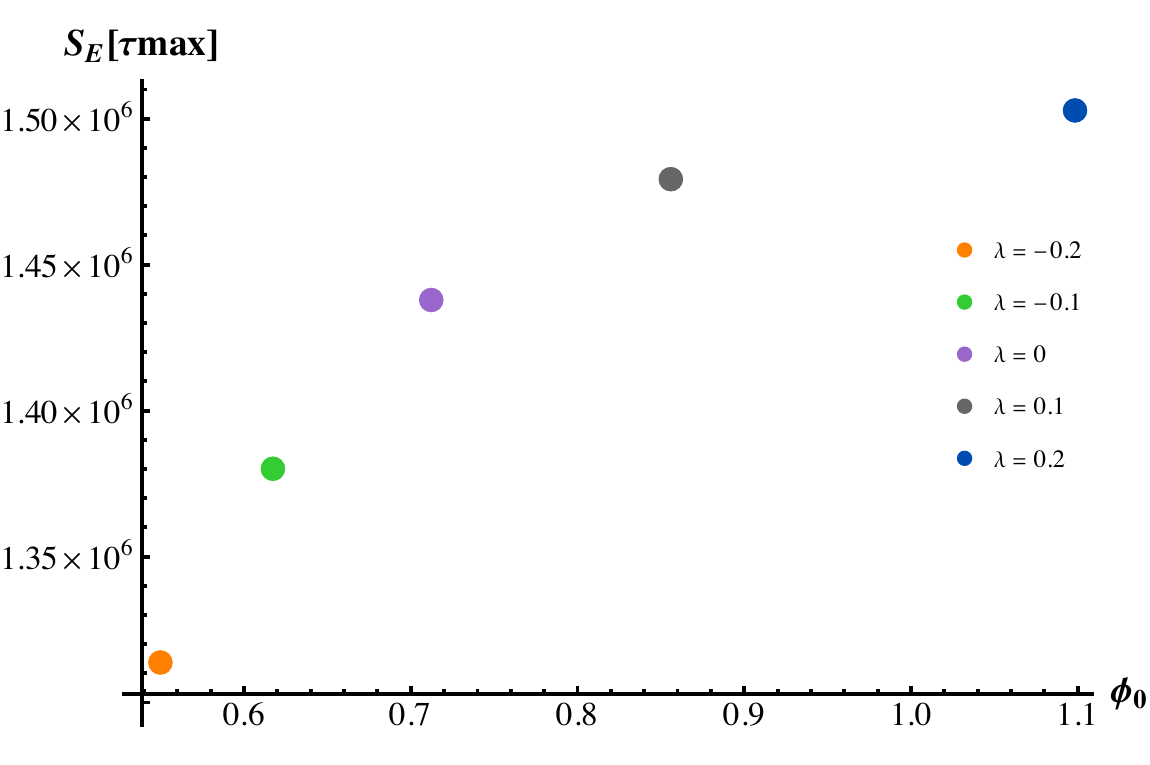}}
	\vspace{-0.4cm}
	\caption{The left panel shows the Euclidean action versus $\tau$ for the solutions in Fig.~\ref{fig:diflammin}, and the right panel shows their asymptotic values. Remarkably, the action demonstrates monotonicity with respect to $\lambda$.}
	\label{fig:diflamminSE}
\end{figure}

More comprehensively, we can investigate the influence of different coupling parameters on the evolution of wormholes. Since each set of parameters corresponds to multiple \(\phi_0\) solutions, this study selects the smallest \(\phi_0\) solution for comparative analysis. Using the shooting method to numerically solve the first and third equations in Eq.~\eqref{e25}, we present the evolution of the scale factor and the dilaton field under different parameters in Fig.~\ref{fig:diflammin}, and show the evolution trajectory and final value of the Euclidean action under the corresponding parameters in Fig.~\ref{fig:diflamminSE}. This study focuses on five cases where \(\lambda\) takes the values \(-0.2\), \(-0.1\), \(0\), \(0.1\), and \(0.2\). Fig.~\ref{fig:diflammin} indicates that the value of \(\phi_0\) increases monotonically with \(\lambda\), while the value of \(a_0\) decreases monotonically with increasing \(\lambda\), suggesting that wormholes with larger coupling parameters have smaller initial throat radii. Furthermore, Fig.~\ref{fig:diflamminSE} depicts the evolutionary trajectory and the final asymptotic value of the Euclidean action for these fundamental solutions. As shown in Fig.~\ref{fig:diflamminSE}, for the fundamental branch with the minimum initial field $\phi_{0,min}$, the Euclidean action increases as the coupling parameter $\lambda$ grows. This indicates that for a simple non-oscillatory evolution, the coupling raises the action. For the oscillatory solutions, however, the coupling parameter $\lambda$ increases the oscillatory behavior of the evolution. This increased oscillation directly results in a lower Euclidean action, as discussed in Fig.~\ref{fig:lam0.1} and Fig.~\ref{fig:lam0.1SE}.

\begin{figure}[htbp]
	\centering
	\subfigure{\includegraphics[width=0.47\textwidth]{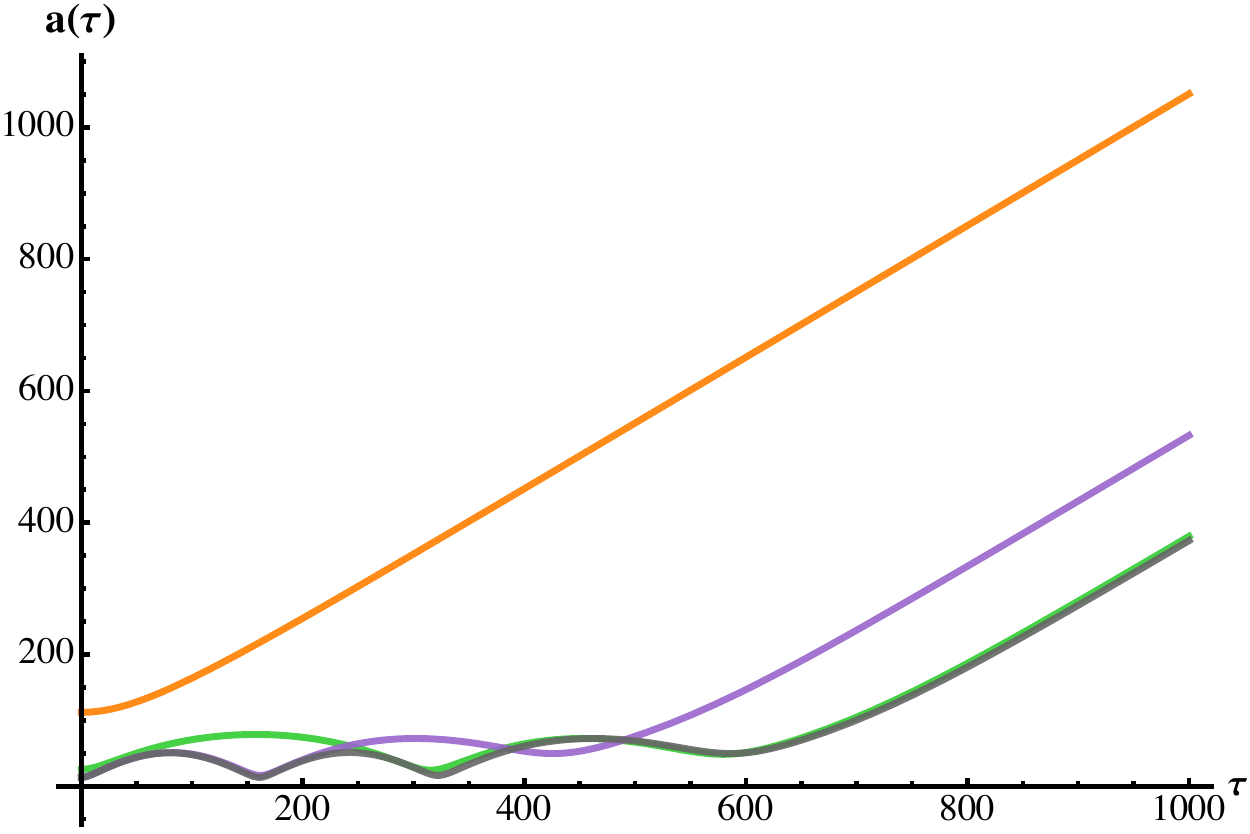}}
	\subfigure{\includegraphics[width=0.47\textwidth]{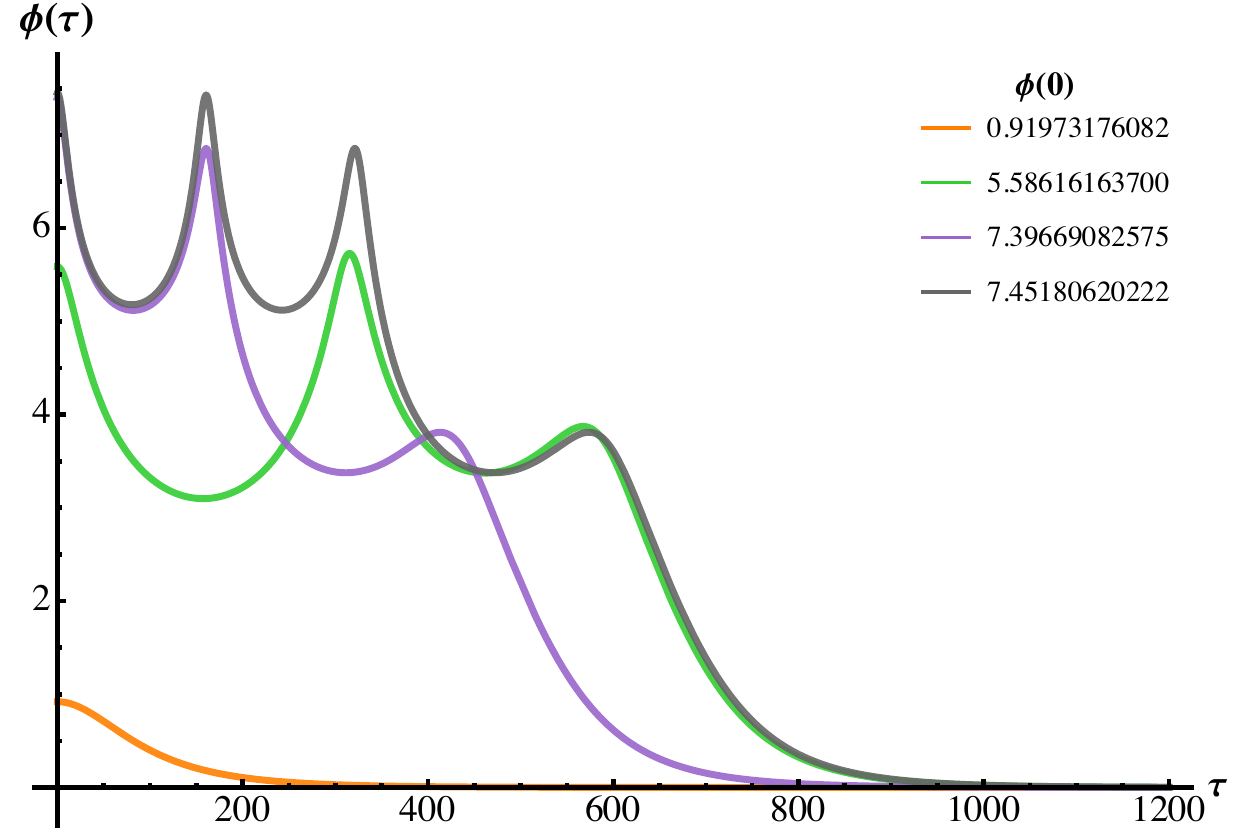}}
	\vspace{-0.4cm}
	\caption{Evolution of the contracting (GS-type) wormhole solutions with modified parameters: $\kappa = 1$, $\beta = 1.3$, $N = 35000$, $m = 0.01$, and $\lambda = 0.1$. The initial values of the dilaton field are $0.91973176082$, $5.58616163700$, $7.39669082575$, and $7.45180620222$, respectively. The oscillatory nature of the scale factor (left) and dilaton field (right) remains robust under parameter variation.}
	\label{fig:lam0.1_new_N35000}
\end{figure}

\begin{figure}[htbp]
	\centering
	\subfigure{\includegraphics[width=0.49\textwidth]{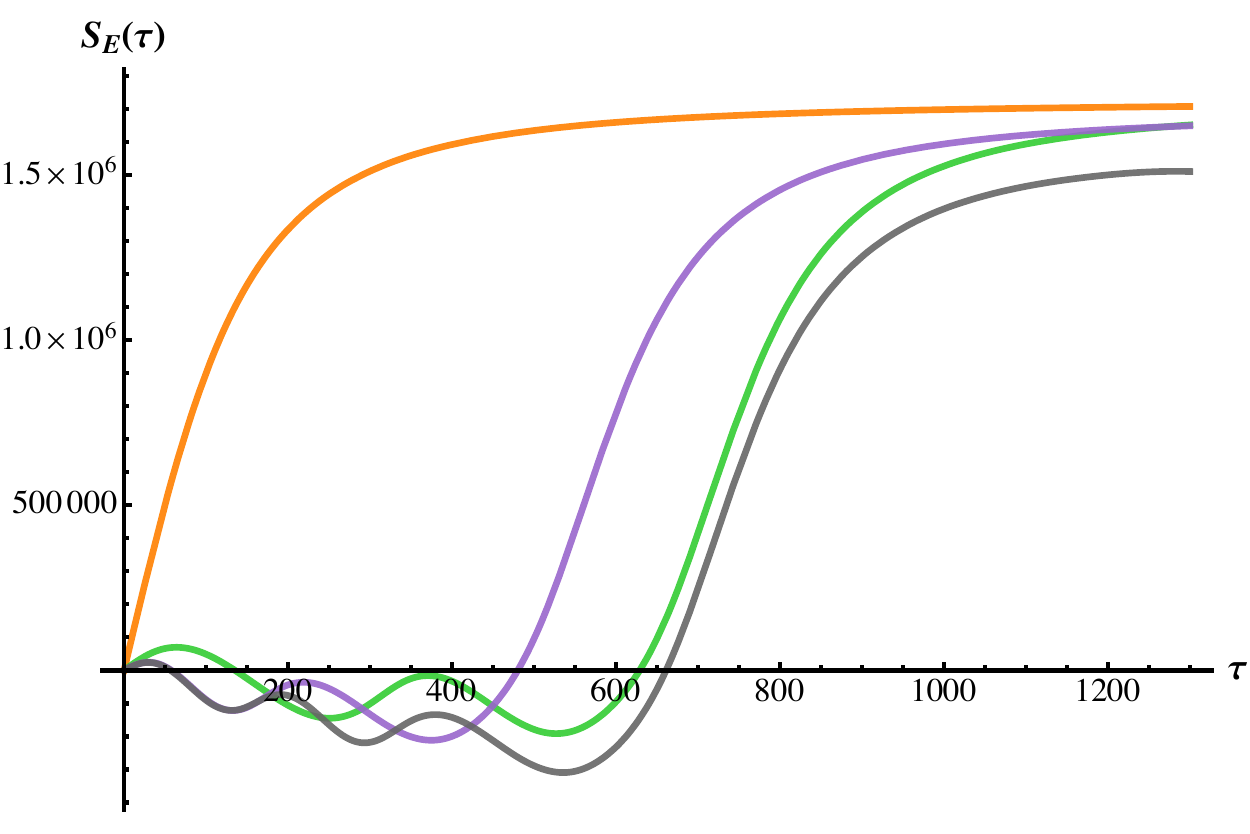}}
	\subfigure{\includegraphics[width=0.49\textwidth]{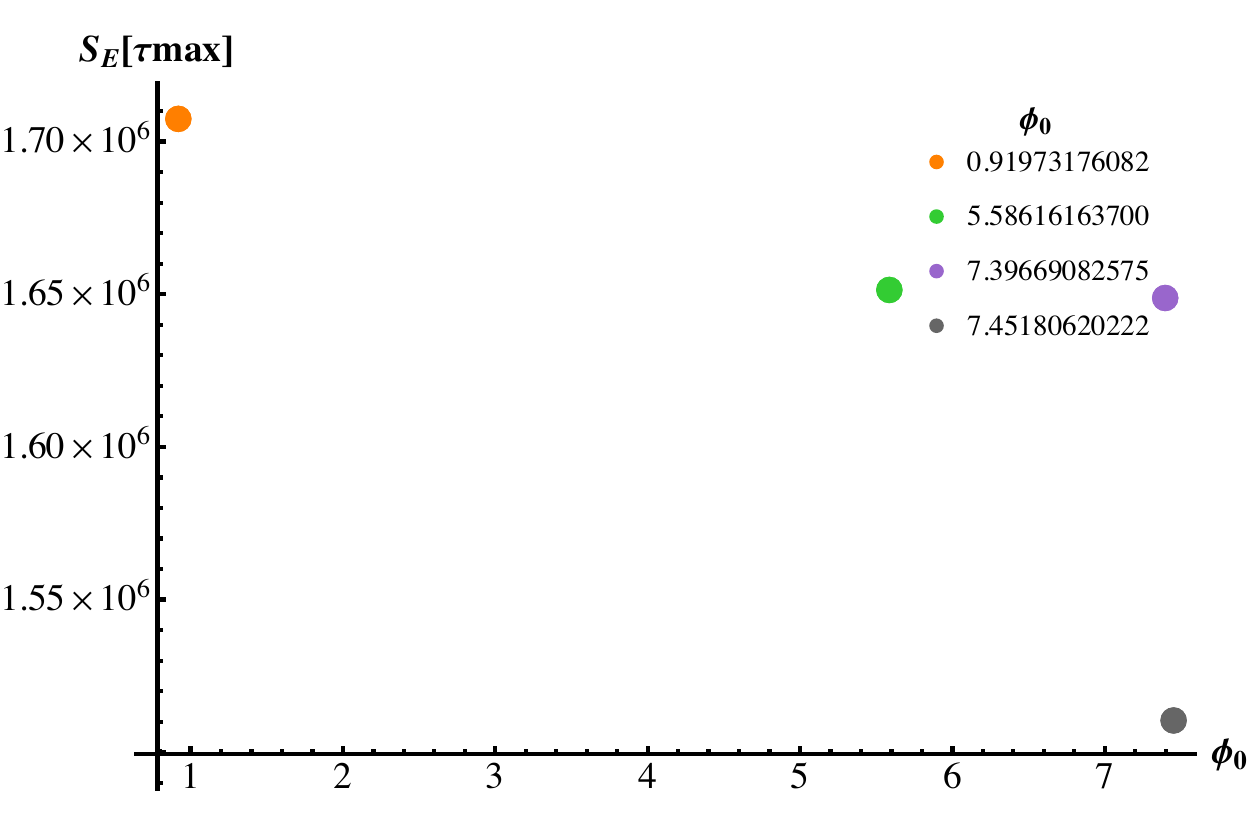}}
	\vspace{-0.4cm}
	\caption{The Euclidean action as a function of $\tau$ (left) and the corresponding asymptotic values (right) for the solutions presented in Fig. \ref{fig:lam0.1_new_N35000}. The overall magnitude of the action increases with $\beta$ and $N$, yet the trend where additional oscillations lead to a decreased action is preserved.}
	\label{fig:lam0.1_new_N35000SE}
\end{figure}

To confirm that the aforementioned mechanisms are not artifacts of our specific parameter choice, we further investigate the system evolution under varied values of the dilatonic coupling $\beta$ and the axion flux $N$. By setting $\beta = 1.3$ and $N = 35000$ while keeping $m = 0.01$ and $\lambda = 0.1$ fixed, we obtain the numerical results are presented in Fig. \ref{fig:lam0.1_new_N35000} and Fig. \ref{fig:lam0.1_new_N35000SE}. The evolution of the scale factor and the dilaton field continues to exhibit characteristic oscillatory dynamics, indicating that these features are robust against variations in the underlying parameters. Although the overall magnitude of the Euclidean action $S_E$ increases with $N$, the non-monotonic dependence of the action on the initial field $\phi_0$ is strictly preserved. These findings confirm that the energetic favorability of complex oscillatory configurations is an intrinsic property of the model rather than an artifact of a specific parameter choice.

Continuing our exploration, we examine an alternative type of wormhole solution within the axion-dilaton modified gravity theory, specifically obtained by evolving the larger root of Eq. $\left(  \ref{e19}\right)$.  Since this wormhole exhibits inflationary expansion in its subsequent Lorentzian evolution, we  refer to it as  \textit{expanding wormhole}~\cite{Jonas:2023ipa}.

\begin{figure}[htbp]
	\centering
	\subfigure{\includegraphics[width=0.47\textwidth]{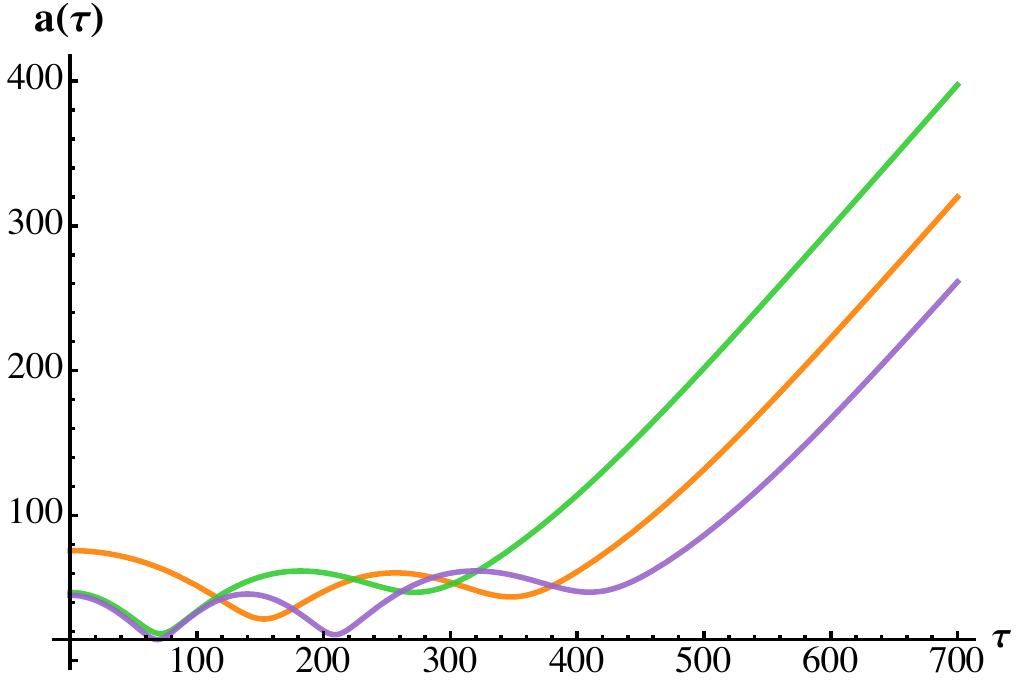}}
	\subfigure{\includegraphics[width=0.47\textwidth]{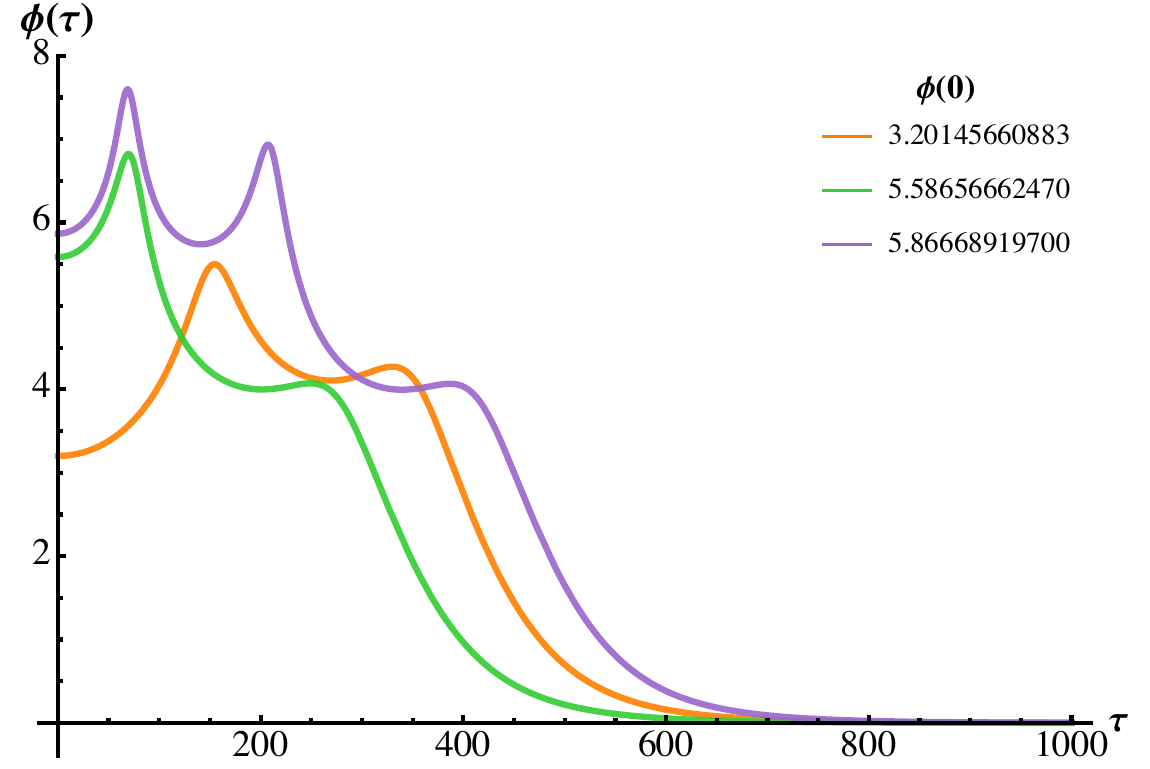}}
	\vspace{-0.4cm}
	\caption{The left panel illustrates the evolution of the scale factor within the expanding wormhole framework, and the right panel shows the evolution of the dilaton field. The depicted solutions correspond to \(\phi_0\) values of $3.20145660883$, $5.58656662470$, and $5.866689196700$, with parameters set at \(\kappa = 1\), \(\beta = 1.2\), \(N = 30000\), \(m = 0.01\), and \(\lambda = 0.1\). Notably, higher \(\phi_0\) values result in more pronounced oscillations in both the scale factor \(a\) and the dilaton field \(\phi\), which significantly impact the wormhole's stability and geometry.}
	
	\label{fig:lam0.1max}
\end{figure}
\begin{figure}[htbp]
	\centering
	\subfigure{\includegraphics[width=0.49\textwidth]{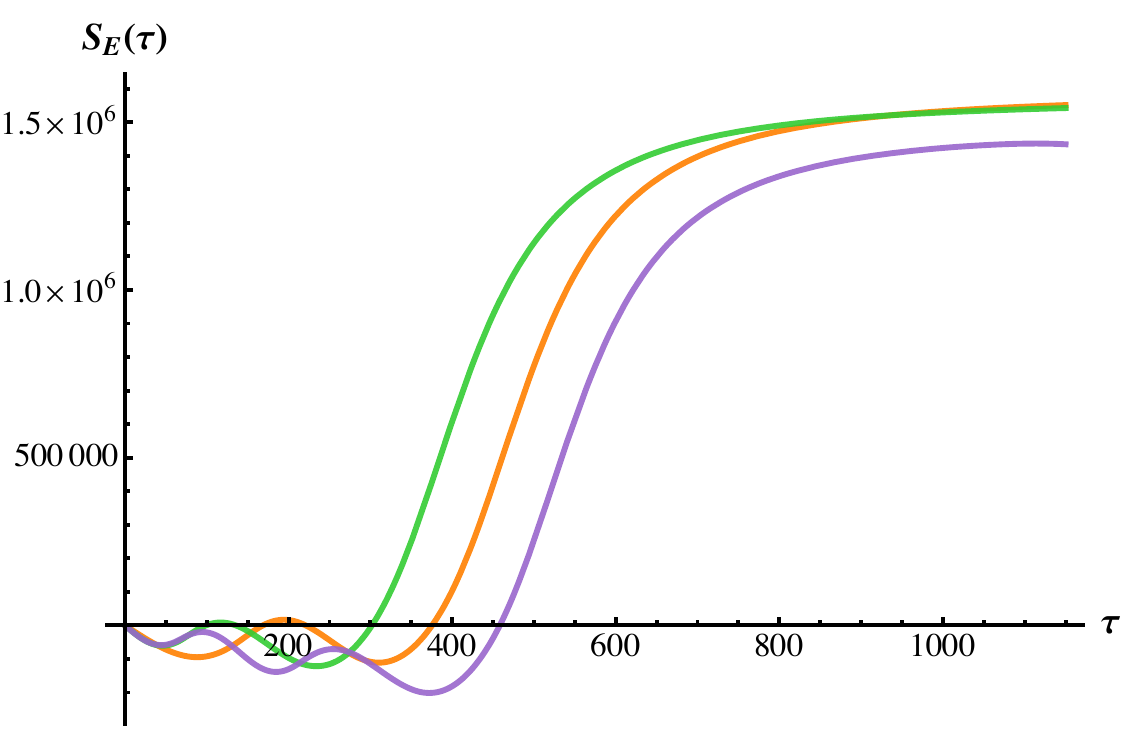}}
	\subfigure{\includegraphics[width=0.49\textwidth]{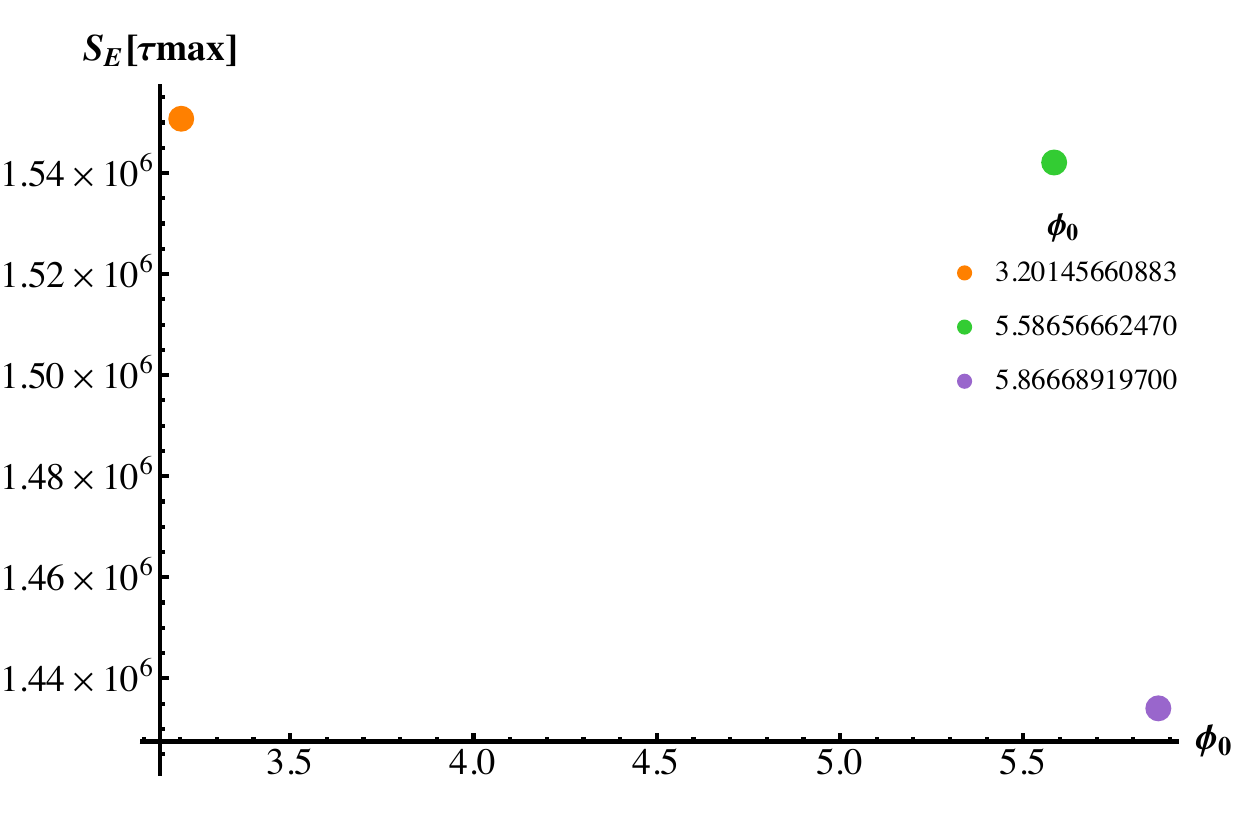}}
	\vspace{-0.4cm}
	\caption{The Euclidean action as a function of 
		$\tau$ (left plot), and the asymptotic values for the solutions in Fig.~\ref{fig:lam0.1max} (right plot). The number of oscillations significantly influences the asymptotic value of the action.}
	\label{fig:lam0.1maxSE}
\end{figure}
The initial set of solutions for expanding wormholes is depicted in Fig.~\ref{fig:lam0.1max}. It is evident that the scale factor exhibits a local maximum at the origin. Similar to the pattern observed for the collapsing wormholes in the previous section, the evolution of the scale factor and the dilaton field becomes increasingly complex with increasing \(\phi_0\), characterized by more oscillatory behavior. Interestingly, the minimum values of the scale factor during oscillations for larger \(\phi_0\) are closer to each other, implying that the sizes of the wormhole throats are more similar.

\begin{figure}[htbp]
	\centering
	\subfigure{\includegraphics[width=0.47\textwidth]{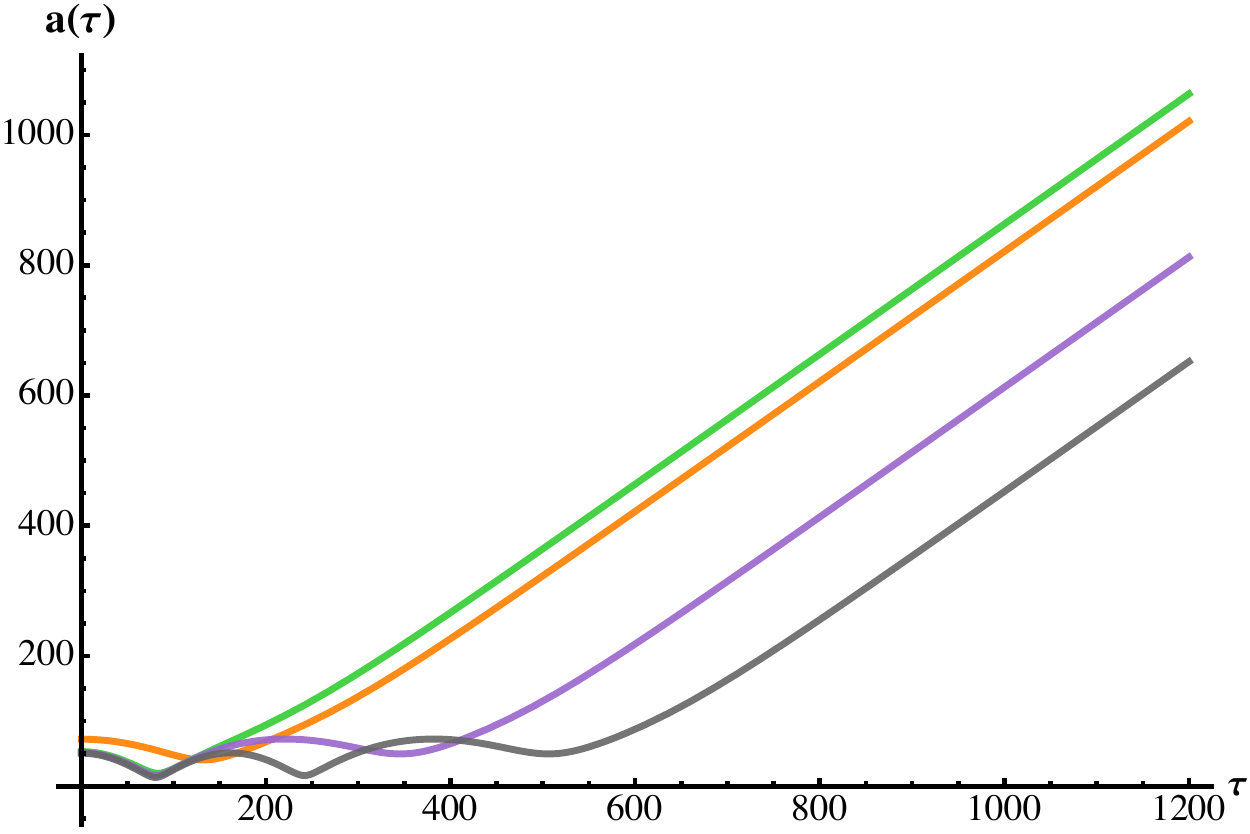}}
	\subfigure{\includegraphics[width=0.47\textwidth]{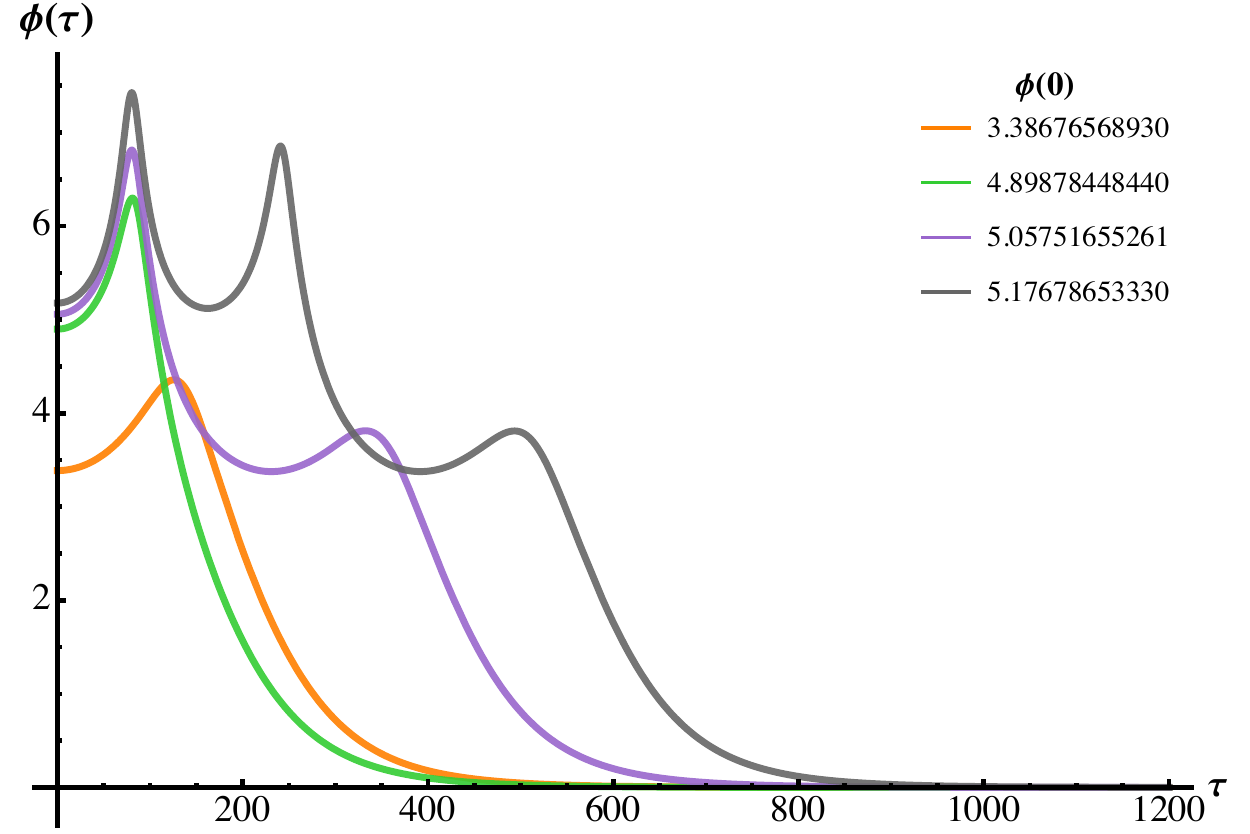}}
	\vspace{-0.4cm}
	\caption{Evolution of the expanding wormhole solutions with parameters: $\kappa = 1$, $\beta = 1.3$, $N = 35000$, $m = 0.01$, and $\lambda = 0.1$. The initial values of the dilaton field are $3.38676568930$, $4.89878448440$, $5.05751655261$, and $5.17678653330$. Complex oscillatory modes continue to manifest at higher parameter values.}
	\label{fig:lam0.1max_new_N35000}
\end{figure}
\begin{figure}[htbp]
	\centering
	\subfigure{\includegraphics[width=0.49\textwidth]{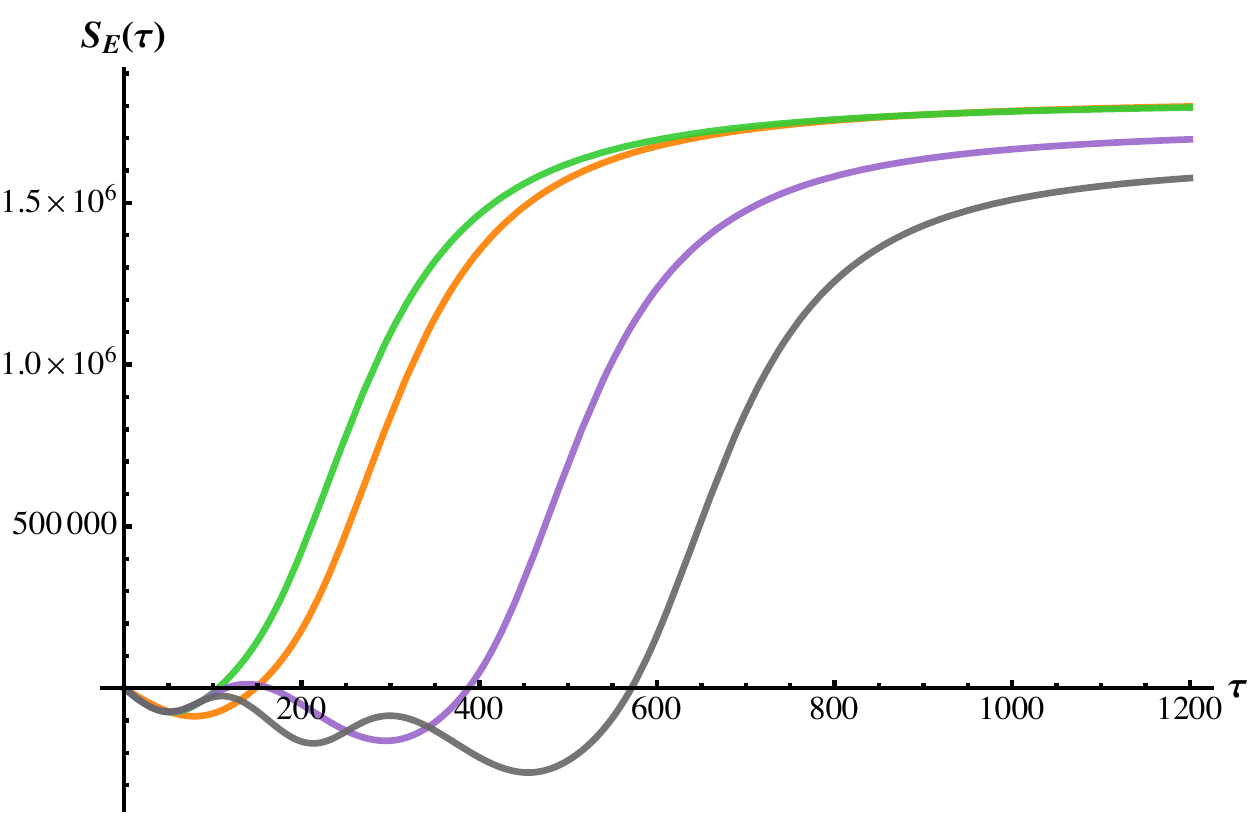}}
	\subfigure{\includegraphics[width=0.49\textwidth]{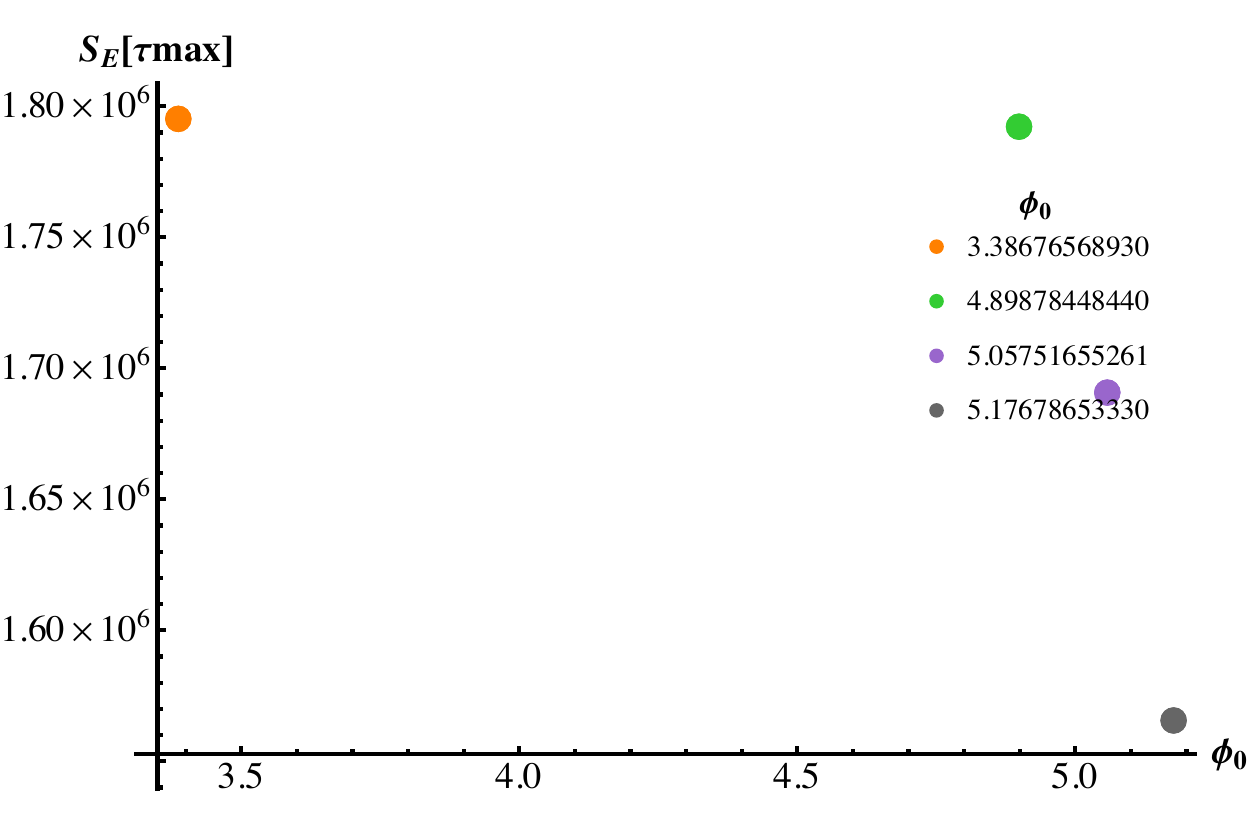}}
	\vspace{-0.4cm}
	\caption{The Euclidean action as a function of $\tau$ (left) and the corresponding asymptotic values (right) for the expanding solutions in Fig.~\ref{fig:lam0.1max_new_N35000}. Consistent with earlier findings, configurations with richer oscillatory features remain probabilistically favored due to their lower action values.}
	\label{fig:lam0.1max_new_N35000SE}
\end{figure}

Regarding the evolution of their action in Fig.~\ref{fig:lam0.1maxSE}, the action decreases immediately as the solutions develop additional inflection points and oscillations, oscillating above and below the x-axis (with more complex solutions lying below the x-axis). As time $\tau$  increases, the action eventually tends towards positive values, consistent with the interpretation of these solutions as mediating tunneling events. This process indicates that the corresponding wormholes will alternately expand and contract, eventually tending towards flatness at large radius.
Figs.~\ref{fig:lam0.1max} and~\ref{fig:lam0.1maxSE}  further demonstrate that the solution represented by the purple curve exhibits more complex evolutionary dynamics and a lower final action value. These results suggest a general trend: solutions with richer oscillatory structures tend to correspond to lower action values, making them more favorable in quantum gravitational processes. Specifically, compared to the green curve, the purple curve corresponds to a solution with more inflection points, more pronounced oscillatory behavior, and more complex dynamical evolution. This enhanced oscillatory character significantly reduces the action of the system, indicating that expanding wormholes with more oscillation modes are probabilistically more likely to exist.

\begin{figure}[htbp]
	\centering
	\subfigure{\includegraphics[width=0.32\textwidth]{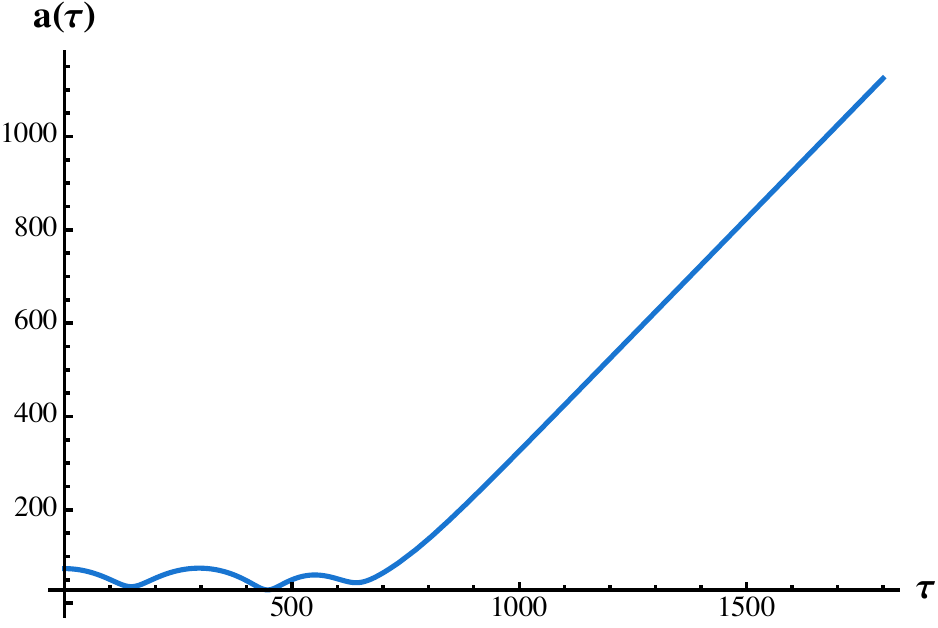}}
	\subfigure{\includegraphics[width=0.32\textwidth]{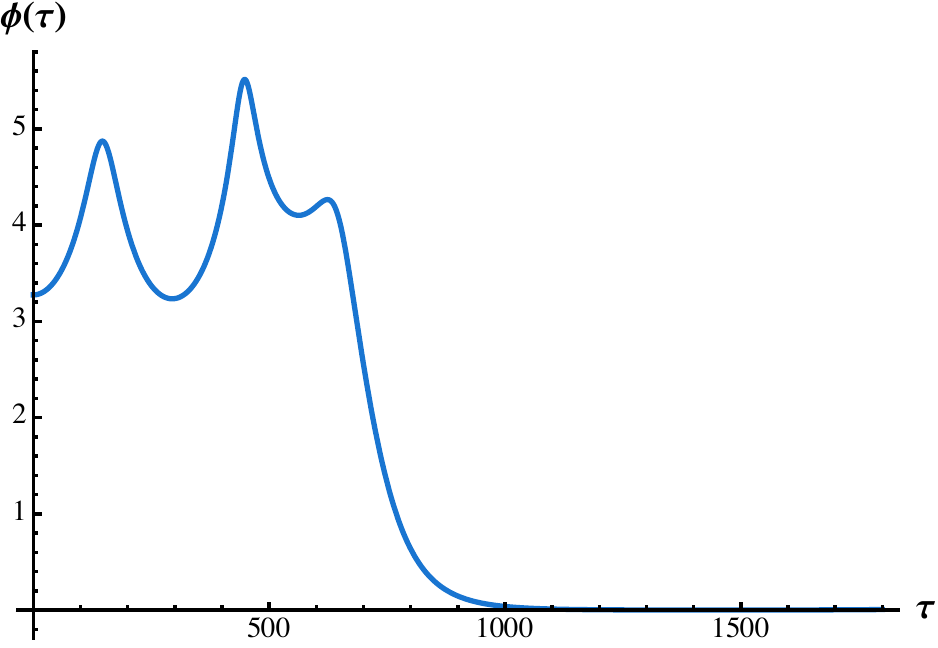}}
	\subfigure{\includegraphics[width=0.32\textwidth]{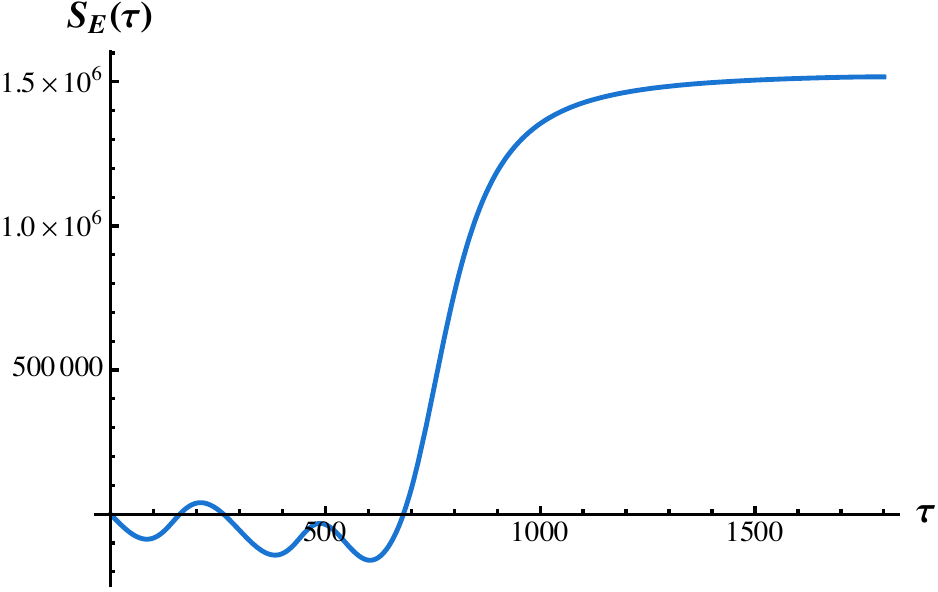}}
	\vspace{-0.4cm}
	\caption{Evolution of an expanding wormhole with \(\phi_0 = 3.27556524203\): the scale factor \(a(\tau)\) (left), dilaton field \(\phi(\tau)\) (middle), and Euclidean action \(S_E\) (right). Notably, despite the relatively small \(\phi_0\), the initial phase of \(\phi(\tau)\) displays a complex pattern of oscillations with increasing amplitudes.
	}
	\label{fig:single_plot}
\end{figure}

Typically, higher values of \(\phi_0\) correlate with more intricate evolutionary processes, as demonstrated in Fig.~\ref{fig:lam0.1max}. Significantly, a special solution with a relatively low \(\phi_0\) value characterized by \(\phi_0= 3.27556524203\) exhibits surprisingly complex oscillatory behavior in the initial phase of the dilaton field's evolution, as illustrated in Fig.~\ref{fig:single_plot}. In contrast to the previously analyzed scenario with \(\phi_0 = 5.86668919670\), where a higher \(\phi_0\) value was expected to result in greater complexity, this particular case, despite its lower \(\phi_0\) value, exhibits a more complex oscillatory pattern in the initial phase of its evolution. Specifically, the amplitude of the dilaton field \(\phi(\tau)\) initially increases and then decreases, taking a longer duration to stabilize at a lower value. This intricate oscillatory behavior during the initial phase suggests that the dynamic evolution of the dilaton field can be highly complex even at relatively lower \(\phi_0\) values, which  contradicts the intuitive hypothesis that the parameter magnitude is the sole determinant of evolutionary complexity, thereby providing novel theoretical insights. Such complexity may have important implications for the stability and geometric configuration of wormhole solutions, offering novel insights into the study of their physical properties.

The dynamic characteristics of the action further indicate that solutions with richer oscillatory features are more probabilistically significant due to their lower action values, suggesting a preference for such configurations in quantum gravitational processes. These results help understand the stability and geometric progression of wormhole solutions. These findings expand the solution space of wormhole solutions in modified gravity theories and provide critical dynamical benchmarks for exploring spacetime topology changes in quantum cosmology.

Similarly, the robustness of the expanding wormhole solutions is verified against parameter variations. As illustrated in Fig.~\ref{fig:lam0.1max_new_N35000} and Fig.~\ref{fig:lam0.1max_new_N35000SE}, adopting the modified parameter set ($\beta = 1.3, N = 35000$) does not qualitatively alter the fundamental characteristics of the expanding branches. Complex oscillatory modes continue to manifest at higher initial field values, and configurations with richer oscillatory features remain probabilistically favored due to their lower action values. This systematic validation across both solution types solidifies the conclusion that the matter-geometry coupling in $F(R,T)$ gravity intrinsically modulates the tunneling probability, ensuring the general applicability of our results to the pre-inflationary epoch.

\section{Inflationary universe  in the EAdS spacetime}\label{sec:4}
In the previous section, we explored axion-dilaton wormhole solutions in an asymptotically flat Euclidean spacetime. That analysis demonstrated that the coupling parameter $\lambda$ plays a crucial role in determining the wormhole's geometry and action. Building on those insights, this section shifts focus to a different physical scenario more directly relevant to cosmology: a ``wineglass'' half-wormhole model within an asymptotically Euclidean Anti-de Sitter (EAdS) spacetime, aiming to address the issue of insufficient inflation duration in the no-boundary proposal.

\begin{figure}[htbp]
	\centering
	\subfigure{\includegraphics[width=0.45\textwidth]{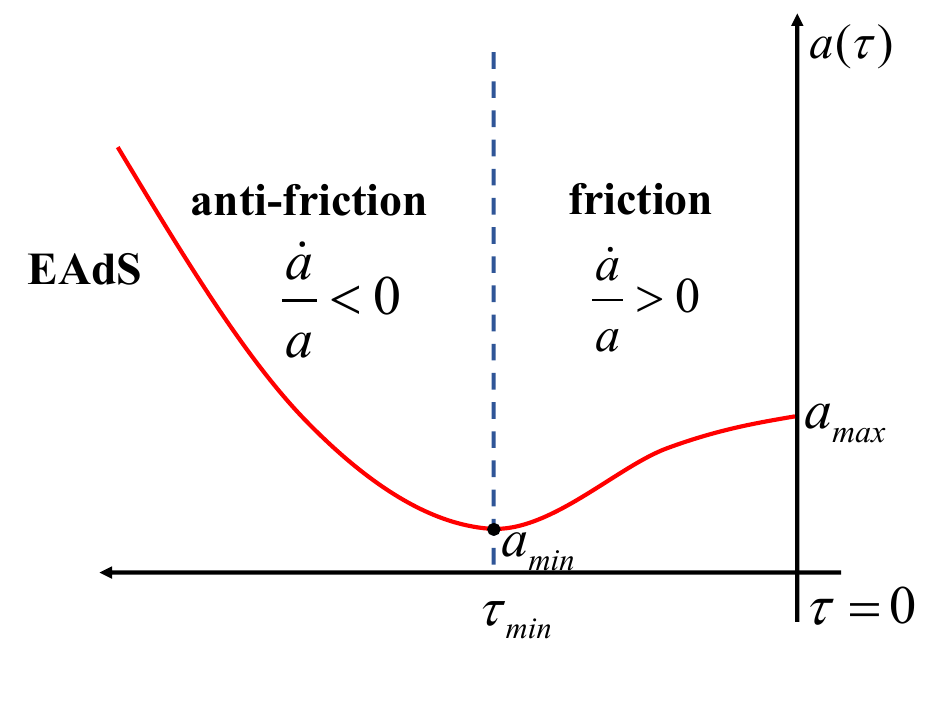}}
	\subfigure{\includegraphics[width=0.45\textwidth]{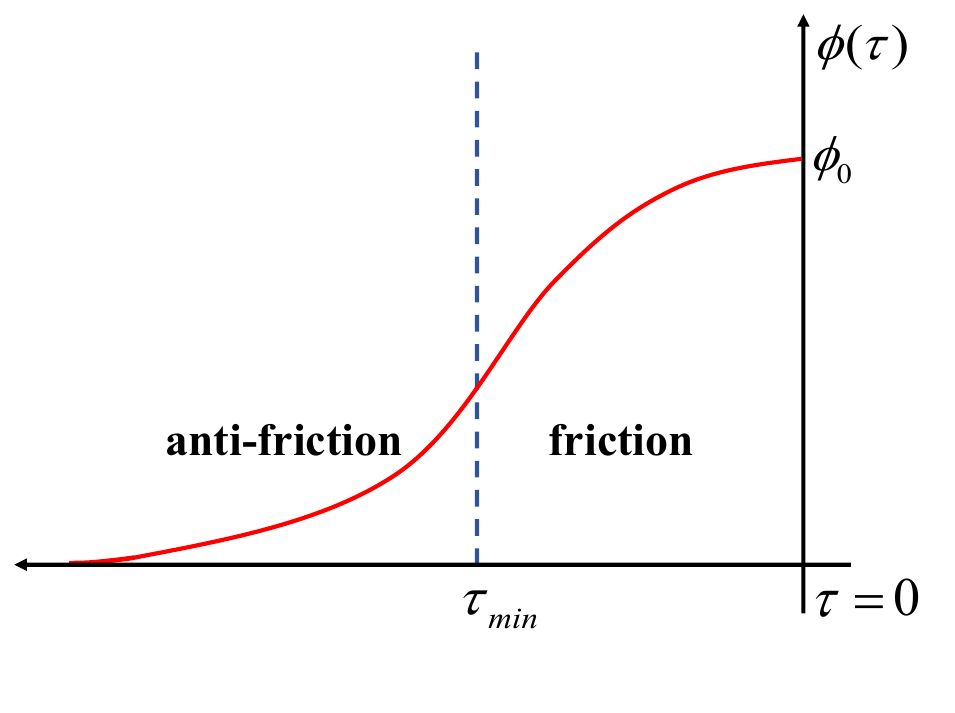}}
	\vspace{-0.4cm}
	\caption{Euclidean evolution of the scale factor \(a(\tau)\) (left) and the scalar field \(\phi(\tau)\) (right) for an \(EAdS\) ``wineglass'' half-wormhole. The evolution is divided into frictional and anti-frictional regions based on the sign of \(\dot{a}/a\).}
	\label{fig:a_EADS}
\end{figure}

\begin{figure}[h]
	\centering
	\includegraphics[width=0.8\textwidth]{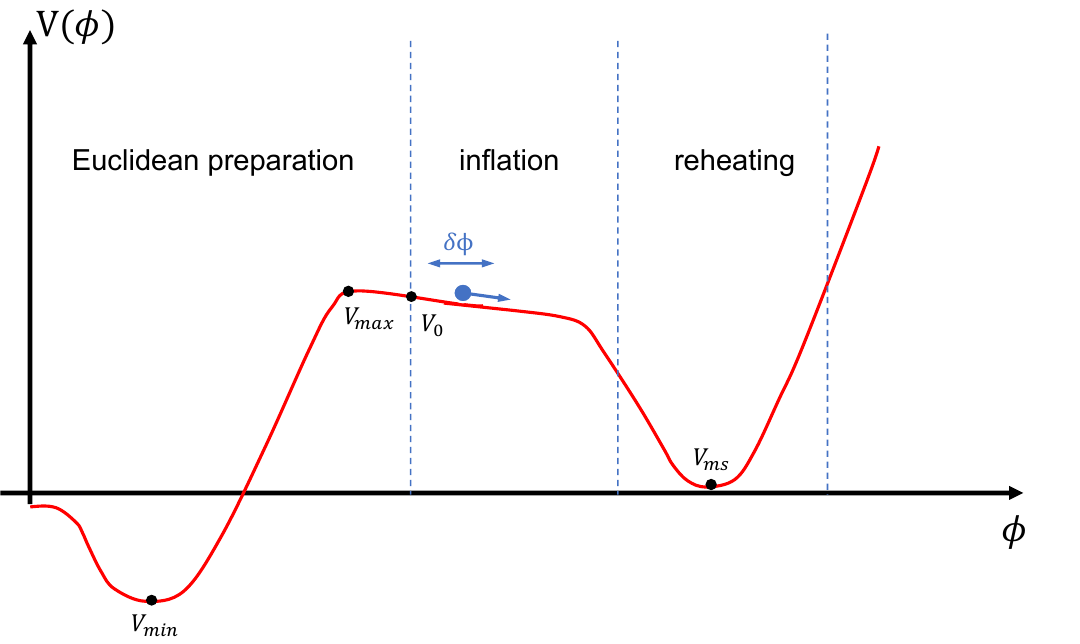}
	\caption{
		The schematic diagram of \( V(\phi) \) corresponds to three physical stages: Euclidean evolution, inflation, and reheating. In the diagram, we have marked a global negative minimum \( \tilde{V}_{\text{min}} \), a positive maximum \( \tilde{V}_{\text{max}} \), and a positive metastable minimum \( \tilde{V}_{\text{ms}} \). }
	\label{fig:slow_roll}
\end{figure}

It is natural to focus on the ``wineglass'' half-wormhole model (featuring an expanding wormhole) within the framework of cosmic expansion in Euclidean AdS spacetime. The model is driven by a scalar field with $\beta=0$, which simplifies the expression to $Q^2=N^2$. Consequently, Eq.~\eqref{e13} reduces to
\begin{equation}
	\begin{aligned}
		2a\ddot{a}+\dot{a}^2-1+\kappa a^2\left((1-\frac{\lambda}{\kappa}) \frac{\dot{\phi}^2}{2}+(1-\frac{2\lambda}{\kappa}) V(\phi)\right)-(1+\frac{\lambda}{\kappa}) \frac{\kappa Q^2}{a^4} &= 0, \\
		\dot{a}^2 - 1 - \frac{\kappa a^2}{3} \left( (1-\frac{\lambda}{\kappa})\frac{\dot{\phi}^2}{2} - (1-\frac{2\lambda}{\kappa})V(\phi) \right) + (1+\frac{\lambda}{\kappa}) \frac{\kappa Q^2}{3a^4} &= 0, \\
		\ddot{\phi} + 3 \frac{\dot{a}}{a} \dot{\phi} - \frac{1 - \frac{2\lambda}{\kappa}}{1 - \frac{\lambda}{\kappa}} \frac{\partial V}{\partial \phi} &= 0.
	\end{aligned}
	\label{e27}
\end{equation}   
The scalar field equation can be viewed as a particle $\phi$ moving in an effective potential $U_E = -V(\phi)$, with a damping term $3\frac{\dot{a}}{a}\dot{\phi}$ whose behavior depends on the sign of $\dot{a}/a$. In the frictional region where $\tau \in (\tau_{\min}, 0)$ and $\dot{a}/a > 0$, this term acts as a friction force that slows down the field evolution. Conversely, in the anti-frictional region where $\tau \in (-\infty, \tau_{\min})$ and $\dot{a}/a < 0$, it becomes an anti-friction term that accelerates the field motion, as illustrated in Fig.~\ref{fig:a_EADS}. This transition occurs at $\tau_{\min}$ where the scale factor reaches its minimum value $a_{\min}$ and the sign of $\dot{a}/a$ changes.
We adjust the boundary conditions such that as \(\tau\) approaches infinity, the solutions asymptotically approach an \(EAdS\) space, given by \(a(\tau) \sim \exp(H_{AdS}|\tau|)\). Additionally, we require that these solutions satisfy the following conditions at \(\tau = 0\),
\begin{equation}
	\ddot{a}(0) < 0, \quad \dot{a}(0) = 0, \quad a(0) = a_{\text{max}}, \quad \dot{\phi}(0) = 0.
\end{equation}
In Eq. $\left(  \ref{e23}\right)$, the value of \(a_{\text{max}}\) remains constrained, indicating that the wormhole throat may extend to the size of the Hubble radius. By maintaining these constraints at \(\tau = 0\), we establish a theoretical foundation for spacetime emergence.  The scalar potential $V(\phi)$ is taken to be of the hilltop type~\cite{Boubekeur:2005zm}, characterized by the qualitative features shown in Fig.~\ref{fig:slow_roll}: a global negative minimum $\tilde{V}_{\min}$, a positive local maximum $\tilde{V}_{\max}$, and a positive metastable minimum $\tilde{V}_{\rm ms}$, with $V \to 0$ asymptotically corresponding to the EAdS vacuum. Our subsequent analytic results in both the thin-wall and thick-wall limits depend only on the local slow-roll expansion of $V$ near $\tilde{\phi}_0$, and are therefore applicable to any potential of this general class. Subsequently, the scalar field evolves within this slow-roll inflationary region, which supports the reheating phase. The validity of the slow-roll approximation is predicated on the assumption of small slow-roll parameters~\cite{Baumann:2009ds},
\begin{equation}
	\epsilon_V \equiv \frac{M_P^2}{16\pi} \left( \frac{V_\phi}{V} \right)^2 \ll 1, \quad \eta_V \equiv \frac{M_P^2 V_{\phi\phi}}{8\pi V} \ll 1,
\end{equation}
corresponding to the potential \( V(\phi) \) in the ``inflation'' region marked in Fig.~\ref{fig:slow_roll}. The number of \( e \)-folds \( N_* \) during inflation is calculated by integrating \( dN \simeq d\phi / M_P \sqrt{\epsilon_V} \) from the horizon to the end of inflation~\cite{Weinberg:2005vy}.
Inflation typically requires between \( O(50) \) and \( O(60) \) \( e \)-folds.

The reduced action is,
\begin{equation}
	\begin{aligned}
		S_E &=  2\pi^{2} \int \mathrm{d}\tau \left[ -\frac{3a\dot{a}^{2}}{\kappa} - \frac{3a}{\kappa} + (1 - \frac{\lambda}{\kappa}) \frac{a^{3}\dot{\phi}^{2}}{2} + (1 - \frac{2\lambda}{\kappa}) a^{3}V +(1 + \frac{\lambda}{\kappa}) \frac{Q^{2}}{a^{3}} \right],\\
		&=2\pi^{2} \int \mathrm{d}\tau \left[ -\frac{3a\dot{a}^{2}}{\kappa} - \frac{3a}{\kappa} +  \frac{ \gamma a^{3}\dot{\phi}^{2}}{2} + \delta a^{3}V +\frac{\alpha Q^{2}}{a^{3}} \right],
	\end{aligned}
\end{equation}
where $\alpha=1+\frac{\lambda}{\kappa}$, $\delta=1-\frac{2\lambda}{\kappa}$, $\gamma=1-\frac{\lambda}{\kappa}.$  Based on the action, we can derive the canonical momenta conjugate to the scale factor \( a \) and the scalar field \( \phi \),
\begin{equation}
	p_a = \frac{\partial \mathcal{L}}{\partial \dot{a}} = -\frac{12\pi^{2}a\dot{a}}{\kappa}, \quad  p_\phi = \frac{\partial \mathcal{L}}{\partial \dot{\phi}} = 2\pi^{2} \gamma a^{3}\dot{\phi}.
\end{equation}
By expressing \( \dot{a} \) and \( \dot{\phi} \) in terms of their respective conjugate momenta, we obtain the Hamiltonian of the system. Due to the invariance under time reparameterization, the resulting Hamiltonian constraint takes the form,
\begin{equation}
	\mathcal{H} =-\frac{\kappa p_a^2}{24\pi^{2}a} + \frac{p_\phi^2}{4\pi^{2} \gamma a^{3}} + 2\pi^{2}\left[ \frac{3a}{\kappa} - \delta a^{3}V -  \frac{\alpha Q^{2}}{a^{3}} \right].
\end{equation}
The classical Hamiltonian constraint is given by \( \mathcal{H} = 0 \). During quantization, the conjugate momenta are replaced by operators~\cite{Jiang:2025asp}.
Substituting these operators into the Hamiltonian constraint yields the Wheeler-DeWitt equation, which is a hyperbolic partial differential equation controlling the quantum behavior of the universe in the minisuperspace model.The introduction of \( A = \log a \) helps resolve operator ordering ambiguities. By further defining \(\tilde{\phi} = \phi / M_{\text{Pl}}\), \(\tilde{V} = \kappa V / 3\), and \(\tilde{Q}^2 = \kappa Q^2 / 3\), the Wheeler-DeWitt equation simplifies to,
\begin{equation}
	\left[ \frac{\partial^2}{\partial A^2} - \frac{1}{\gamma}\frac{\partial^2}{\partial \tilde{\phi}^2} +  \frac{144\pi^{4}}{\kappa^2}  \left(  \delta e^{6A}\tilde{V}(\tilde{\phi}) \right. \right.
	\left. \left. - e^{4A} + \alpha \tilde{Q}^2 \right) \right] \Psi(A, \tilde{\phi}) = 0,	
\end{equation}
where \(\Psi(A, \tilde{\phi})\) is the wave function of the universe~\cite{Hartle:1983ai}, it is restricted by two prominent proposals: the Hartle-Hawking no-boundary (NB) proposal~\cite{Hartle:2007gi,Lehners:2023yrj} and the Vilenkin tunneling  proposal~\cite{Vilenkin:1983xq,Vilenkin:1982de,Jiang:2025asp}. The Hartle-Hawking no-boundary proposal defines the wave function through Euclidean path integrals over compact geometries without boundaries, effectively avoiding singularity issues associated with the Lorentzian Big Bang. This proposal aligns well with the observed simplicity, homogeneity, and isotropy of the early universe, and predicts an approximately Gaussian primordial perturbation spectrum~\cite{Lehners:2023yrj,Maldacena:2024uhs}.
However, the theory predicts insufficient e-foldings during inflation. We attempt to address this issue within the framework of a semi-wormhole model using $F(R,T)$ theory.

The expression for the Euclidean on-shell action at the semi-classical level is given by~\cite{Betzios:2024oli},
\begin{equation}
	\begin{aligned}
		S_E^{\text{on-shell}} &=4\pi^2 \int \mathrm{d}\tau \left[ \frac{2Q^2}{a^3} - a^3 V + \frac{2\lambda}{\kappa} ( a^3 V + \frac{Q^2}{a^3}) \right] + S_{GHY} + S_{c.t.},\\
		&=4\pi^2 \int \mathrm{d}\tau \left( \frac{2\alpha Q^2}{a^3} -  a^3 \delta V \right) + S_{GHY} + S_{c.t.},
	\end{aligned}
\end{equation}
The expression includes the Gibbons-Hawking-York term \(S_{GHY}\) as well as the boundary counterterms $S_{c.t.}$, which are essential for carrying out holographic renormalization, ensuring that the action remains finite in spaces with an asymptotically $EAdS$ boundary. The action consists of two parts corresponding to the frictional and anti-frictional regions. The first part includes contributions from the Euclidean AdS boundary term. As demonstrated in previous studies, the Euclidean AdS action with an $S^3$ boundary makes a positive contribution to the action, regardless of the initial value $V(\phi_0)$ of the inflation potential~\cite{Jafferis:2011zi,Taylor:2016kic,Ghosh:2018qtg}. Thus, this part is considered a positive constant. Our analysis then focuses on the integral of the second part. This discussion is further divided into two cases: \(a_{\min} \ll a_{\max}\) and \(a_{\min} \approx a_{\max}\). Analytically, these regimes are characterized by distinct parameter constraints: the thin-wall limit is defined by $a_{\min}\sqrt{\delta\tilde{V}_0} \ll 1$, while the thick-wall limit is governed by the relation $\bar{a} = r/\sqrt{\delta\tilde{V}_0}$, $r \sim \mathcal{O}(1)$. Here, $\bar{a}$ represents the characteristic scale factor of the wormhole throat, a quantity that will be formally defined and analyzed in the subsequent section. A rigorous algebraic derivation of these conditions is provided in \ref{app1}.

\subsection{The case of \(a_{\min} \ll a_{\max}\)}
As discussed in previous studies~\cite{Betzios:2024oli,Coleman:1980aw}, when \(\tau\) approaches \(\tau_{\min}\), a very narrow interval emerges where the time derivative of the scale factor $\dot{a}$ approaches zero. This interval defines a ``thin-wall'' transition zone, in which the scalar field undergoes a rapid transition from \(\tilde{\phi}_{\tau_{\min}}\) to  \(\tilde{\phi}_0\). During this transition, the scale factor stabilizes at a value close to \(a_{\min}\),  remaining nearly constant. Within this ``thin-wall'' region, the action is
{\small \begin{equation}
		S_E^{\text{thin}} =  \frac{6\pi^2}{\kappa} \int_{\text{thin}} \mathrm{d}\tau \left(\frac{2\alpha\tilde{Q}^2}{a_{\min}^3} - a_{\min}^3 \delta \tilde{V}  \right)
		\simeq  \frac{12\alpha\pi^2 \tilde{Q}^2}{\kappa a_{\text{min}}^3} \Delta \tau_{\text{thin}}.
\end{equation}}
This region is characterized by a larger temporal interval $\Delta \tau$ and $\dot{a}$ can not be neglected. Within this region, the potential energy stabilizes at a constant value, causing the scale factor to enter a ``slow roll'' phase that continues until the scale factor reaches its maximum value $a_{\max}$. The corresponding action is given by
\begin{equation}
	S_E^{\text{thick}} =\frac{12\pi^2}{\kappa} \int_{\text{thick}} \mathrm{d}\tau \left(\frac{2\alpha\tilde{Q}^2}{a^3} - a^3 \delta\tilde{V} ) \right).
\end{equation}
It is convenient to change the integration variable from time $\tau$ to the scale factor $a$ by utilizing the Friedmann constraint (the second expression in Eq. $\left(  \ref{e27}\right)$),
\begin{equation}
	S_E^{\text{thick}} = \frac{12\pi^2}{\kappa} \int_{a_{\min}}^{a_{\max}} \frac{\left( \frac{2\alpha\tilde{Q}^2}{a^3} - a^3\delta \tilde{V}\right)}{\sqrt{1 - a^2 \delta  \tilde{V} - \frac{\alpha\tilde{Q}^2}{a^4}}}da.
\end{equation}
Given the assumption that \(\tilde{\phi} \simeq \tilde{\phi}_0\) is approximately constant, the value of \(\tilde{V}\) can be determined accordingly. According to Eq. $\left(  \ref{e22}\right)$, the condition \(a_{\text{min}} \ll a_{\text{max}}\) is satisfied only when $\alpha\tilde{Q} \to 0$ and  \(\lambda\) is constrained to the range \(\lambda < \kappa/2\) (which means $\delta>0$). Under these conditions, \(a_{\text{max}}\) can be expressed as \(a_{\text{max}} = \frac{1}{\sqrt{\delta \tilde{V}(\phi_0)}}\). Therefore, the integral can be calculated accordingly,
\begin{equation}
	\begin{aligned}
		S_E^{\text{thick}} \simeq & \frac{12\pi^2 \alpha\tilde{Q}^2}{\kappa} \left[ \frac{\sqrt{1 - a_{\min}^2 \delta\tilde{V}_0}}{a_{\min}^2} + \delta\tilde{V}_0 \tanh^{-1}\sqrt{1 - a_{\min}^2 \delta\tilde{V}_0} \right] \\
		& - \frac{4\pi^2}{\kappa\alpha \tilde{V}_0} \left( 1 - a_{\min}^2 \delta\tilde{V}_0 \right)^{3/2}.
	\end{aligned}
\end{equation}
Due to the condition \(a_{\text{min}} \sqrt{\delta\tilde{V}_0} \ll 1\), we can expand the preceding equation,
\begin{equation}
	S_E^{\text{thick}} \simeq  \frac{12\pi^2 \alpha\tilde{Q}^2}{\kappa} \left[ \frac{1}{a_{min}^2} \right. \left. - \delta\tilde{V}_0 \log \frac{a_{min} \sqrt{\delta\tilde{V}_0}}{2} \right] 
	- \frac{4\pi^2}{\kappa\delta\tilde{V}_0} + O(a_{min}^2 \delta\tilde{V}_0).
\end{equation}
Under the condition of $\alpha\tilde{Q} = 0$, we return to the theoretical framework proposed in reference \cite{Betzios:2024oli}. Within this framework, the Euclidean space is characterized by a smooth half \(S^4\) geometry, and the \(EAdS\) asymptotic region becomes completely detached. As a result, the action is reduced to $-{4\pi^2}/{\kappa \tilde{V}_0}$, which is exactly half of the action of a de Sitter instanton, consistent with the no-boundary proposal. However, it is crucial to examine the cases where $\alpha\tilde{Q} $ is small but non-zero, especially in the framework of \(F(R,T)\) theory. To further explore these scenarios, we proceed to differentiate the previously discussed results with respect to \( \tilde{V_0}\),
\begin{equation}
	\frac{\partial S_E^{\text{thick}}}{\partial \tilde{V}_0}  = \frac{4\pi^2}{\kappa\delta\tilde{V}_0^2} -\frac{12\pi^2\alpha\delta\tilde{Q}^2}{\kappa} 
	\left[\frac{1}{2}+ \log \frac{a_{min} \sqrt{\delta\tilde{V}_0}}{2} \right] .
\end{equation}
To gain a deeper understanding of the properties at the extremum points, it is necessary to compute the second derivative of \( \tilde{V_0}\).
\begin{equation}
	\frac{\partial ^2 S_E^{\text{thick}}}{\partial \tilde{V}_0 ^2}=\frac{-6\pi^2\alpha\delta^2\tilde{Q}^2\tilde{V_0}^2-8\pi^2}{\kappa\delta\tilde{V_0}^3}.
\end{equation}
When $\alpha\tilde{Q} $ is small but non-zero, the action \(S_E\) has an unstable maximum at \(\tilde{V}_*\), where the corresponding probability \(P = e^{-S_E}\) reaches a local minimum. In this model, \(\tilde{V}_0\) is constrained within the range \(\tilde{V}_{\text{ms}} \leq \tilde{V}_0 \leq \tilde{V}_{\text{max}}\) (between the metastable minimum and the local maximum, as shown in Fig.~\ref{fig:slow_roll}). When \(\tilde{V}_0 < \tilde{V}_*\), the action \(S_E\) increases with \(\tilde{V}_0\), leading to a decrease in probability \(P\) toward \(\tilde{V}_{\text{ms}}\). Conversely, when \(\tilde{V}_0 > \tilde{V}_*\), the action \(S_E\) decreases with \(\tilde{V}_0\), resulting in an increase in probability \(P\) toward \(\tilde{V}_{\text{max}}\). Since \(\tilde{V}_*\) is an unstable point, the system is more likely to reside at the boundary extrema with higher probability. When \(\tilde{V}_0\) approaches \(\tilde{V}_{\text{max}}\), the scalar field enters the ``hilltop'' slow-roll regime, allowing for a longer duration of inflation. This produces sufficient e-folds to address the issue of insufficient inflationary duration in the no-boundary proposal.

For investigating the impact of \(\lambda\) on the action, we assume that \(Q\) is non-zero and small. The parameter \(\lambda\) satisfies the original basic constraint \(\lambda < \kappa/2\) (implying $\delta>0$) and simultaneously fulfills the condition $\alpha \tilde{Q} \to 0$,
\begin{equation}
	\begin{split}
		\frac{\partial S_{E}^{\text{thick}}}{\partial \lambda} = & \frac{12\pi^2\tilde{Q}^2}{\kappa^2 a_{min}^2} + \frac{12\pi^2\alpha\tilde{Q}^2 \tilde{V}_0}{\kappa^2} - \frac{8\pi^2}{\kappa^2 \delta^2 \tilde{V}_0}+ \frac{24\pi^2 \lambda \tilde{Q}^2 \tilde{V}_0}{\kappa^3} \log \frac{a_{min}\sqrt{\delta\tilde{V}_0}}{2}.
	\end{split}
\end{equation}
And the second-order derivative is,
\begin{equation}
	\frac{\partial^2 S_E^{\text{thick}}}{\partial \lambda^2} =  \frac{12\pi^2 \tilde{Q}^2 \tilde{V}_0}{\kappa^3} - \frac{32\pi^2}{\kappa^3 \delta^3 \tilde{V}_0}- \frac{24\pi^2 \lambda \tilde{Q}^2 \tilde{V}_0}{\kappa^4 \delta}  + \frac{24\pi^2 \tilde{Q}^2 \tilde{V}_0}{\kappa^3} \log \frac{a_{min} \sqrt{\delta \tilde{V}_0}}{2}.
	\label{e45}
\end{equation}
We assume that there exists a specific value of the coupling parameter \( \lambda_* \) such that the partial derivative of the action with respect to \( \lambda \) vanishes at this point, $\left.\frac{\partial S_{E}^{\text{thick}}}{\partial \lambda}\right|_{\lambda =\lambda_*} = 0.$ This indicates that at \( \lambda = \lambda_* \), the action \( S_{E}^{\text{thick-wall}} \) has an extremum with respect to \( \lambda \). To further investigate the significance of this extremum condition, we substitute \( \lambda_* \) into Eq. $\left(  \ref{e45}\right)$  for simplification, which yields

\begin{equation}
	\left.\frac{\partial^2 S_E^{\text{thick}}}{\partial \lambda^2}\right|_{\lambda=\lambda_*} =  	\frac{8\pi^2}{\kappa^3\delta^3 \tilde{V}_0} (\kappa - 6\lambda_*) 
	- \frac{12\pi^2 \tilde{Q}^2 \tilde{V}_0}{\kappa^2 \lambda_*}  
	- \frac{12\pi^2 \tilde{Q}^2}{\kappa^2 a_{\min}^2\lambda_*}.
\end{equation}

Now, we conduct a detailed discussion on different value ranges of the parameter \(\lambda_*\). We find that when \(\frac{\kappa}{6}<\lambda_*<\frac{\kappa}{2}\), the second - derivative is negative, indicating that the action reaches a maximum in this interval, while the probability weight \(P\) reaches a minimum. This is unfavorable for this parameter configuration. When \(0<\lambda_*<\frac{\kappa}{6}\),  due to the uncertainty of the parameters, it is difficult to determine the sign of the second - derivative of the action with respect to \(\lambda\). When \(\lambda_* < 0\), the second - derivative is positive, and the action reaches a minimum, while the probability weight \(P\) reaches a maximum. This indicates that a negative value for the coupling parameter $\lambda$ is physically favored. Therefore, we refine the constraint from  $\lambda<\frac{\kappa}{2}$ to $\lambda<0$. This ensures the probability distribution peaks at large values of the potential  $\tilde{V_0}$, which favors a prolonged period of inflation and resolves the issue of the standard no-boundary proposal predicting short-lived inflation.

Consequently, we first identify that the baseline constraint $\lambda < \kappa/2$ (ensuring $\delta > 0$) is essential for the physical consistency of the scale factor. Within this thin-wall limit ($a_{min} \ll a_{max}$), we find that restricting the coupling parameter to $\lambda < 0$ effectively minimizes the Euclidean action, thereby favoring initial states that lead to prolonged inflation. This establishes a preliminary viable range for the coupling, which will be further refined in the context of the thick-wall regime.

\subsection{The case of  \(a_{\min} \approx a_{\max}\)}

In this case, the region \(\dot{a}/a \approx 0\) ( \(\dot{a} = 0\)) is broad, corresponding to a thick-wall region. The region where $\dot{a} \neq 0$ has now become significantly compressed, forming a thin-wall region. Since the potential is approximately constant in this thin region, the integral contribution is independent of \(\tilde{V}_0\) and can be neglected in the first-order approximation. 

In the thick-wall region, where \(\bar{a}\) denotes the average value between \(a_{\text{min}}\) and \(a_{\text{max}}\), it is expressed as \(\bar{a} = \frac{r}{\sqrt{\delta\tilde{V}_0}}\) with \(r \sim \mathcal{O} (1)\). The action is given by
\begin{equation}
	S_E^{\text{thick}} \simeq  \frac{12\pi^2}{\kappa} \int_{\text{thick}} d\tau \left[ \frac{2\alpha\tilde{Q}^2}{\bar{a}^3} -\delta \bar{a}^3 \tilde{V}(\tilde{\phi}) \right].
\end{equation}
Integrating the third expression of Eq. $\left(  \ref{e27}\right)$  yields,
\begin{equation}
	\frac{1}{6} \dot{\tilde{\phi}}^2- W(\tau) =\frac{\delta}{\gamma}( \tilde{V}(\tilde{\phi})  \ -\tilde{V}_{\tau=-\infty}) ,
\end{equation}
$W(\tau) = - \int_{-\infty}^{\tau} d{\tau} \frac{\dot{a}} {a}\dot{\tilde{\phi}}^2$ represents the total work of friction, which is divided into positive and negative contributions in the regions of anti-friction and friction, respectively, given by \( W_{\text{friction}}^{\text{total}} = W_{\text{anti-friction}}^+ + W_{\text{friction}}^- \). Consequently, the integral of the action can be expressed as follows,
\begin{equation}		
	S_E^{\text{thick}} \simeq  \frac{12\pi^2}{\kappa} \int_{\tilde{\phi}_{\tau_{\text{min}}}}^{\tilde{\phi}_0}  \frac{d\tilde{\phi}}{\sqrt{\frac{6\delta}{\gamma}\tilde{V}(\tilde{\phi}) - C)}}\left[ {2\alpha\frac{\tilde{Q}^2}{\bar{a}^3} -\delta \bar{a}^3 \tilde{V}(\tilde{\phi})} \right],
\end{equation}
where \( C \) is a constant satisfying \( V_{\text{min}} < C < V_{\tau_{\text{min}}} \). Within the thick-walled region where \(\dot{a}\approx 0\), the scalar field evolves from \(\tilde{\phi}_{\tau_{min}}\) and approaches \(\tilde{\phi}_0\). During this evolution, the potential energy can be approximated by a Taylor expansion around \(\tilde{\phi}_0\), given by \(\tilde{V}(\tilde{\phi}) = \tilde{V}_0(1 - \epsilon_{\tilde{V}} \tilde{\phi})\), where \(\epsilon_{\tilde{V}} \ll 1\) is associated with the slow-roll parameter. Then the above equation can be expressed as,
\begin{equation}
	S_{E}^{\text{thick}} \approx  \sqrt{\frac{2}{3}}  \frac{4\pi^{2} \sqrt{\frac{\delta}{\gamma}(\tilde{V}_{0} - C)} }{\kappa \bar{a}^{3} \epsilon_{\tilde{V}} \tilde{V}_{0}}  \left[ -6 \alpha \tilde{Q}^{2} + \bar{a}^{6} \delta (2C + \tilde{V}_{0}) \right].
\end{equation}
Substitute \(\bar{a} = \frac{r}{\sqrt{\delta\tilde{V}_0}}\) into the expression and apply the condition \(\left.\frac{\partial S_{E}^{\text{thick}}}{\partial \tilde{V}_0}\right|_{\tilde{V}_0=\tilde{V}_0^*} = 0\) to \(\left.\frac{\partial^2 S_{E}^{\text{thick}}}{\partial \tilde{V}_0^2}\right|_{\tilde{V}_0=\tilde{V}_0^*}\) to assess the nature of the extremum point \(\tilde{V}_0^*\),
\begin{equation}
	\begin{aligned}
		\frac{\partial ^2 S_E^{\text{thick}}}{\partial \tilde{V}_0 ^2} = & A \left[ 70 C^3 r^6 \kappa^3 + 24 C r^6 \tilde{V}_0^2 \kappa^3 \right. \\
		& \left. + 8 r^6 \tilde{V}_0^3 \kappa^3 + 3 C^2 \tilde{V}_0 \left( -35 r^6 \kappa^3 + 2 \tilde{Q}^2 \tilde{V}_0^2 \kappa^3 \delta^2 \alpha \right) \right],
	\end{aligned}
	\label{eq:47}
\end{equation}
where
\begin{equation}
	A=\frac{\sqrt{2} \pi ^2  }{\kappa ^4 r^3 \tilde{V}_0^4\epsilon_{\tilde{V}}   \sqrt{3\gamma\tilde{V}_0} (\tilde{V}_0-C)^{3/2}}>0.
\end{equation}
By imposing the condition $\lambda < -\kappa$, we ensure that the factor $\alpha$ becomes negative.  This makes the \(\tilde{Q}^2\) term a negative contribution, making it possible for the entire expression to be negative and thus creating the desired unstable maximum in the action. Similarly, if the minimum value of the inflation potential, \(\tilde{V}_{ms}\), is greater than \(\tilde{V}_*\), the value of \(S_E\) will diminish as \(\tilde{V}_0\) grows larger. Consequently, this reduction in \(S_E\) increases the likelihood of creating a universe that undergoes a prolonged period of inflation. 

Next, we examine the impact of \(\lambda\) on the action,
\begin{equation}
	\frac{\partial S_{E}^{\text{thick}}}{\partial \lambda} = B \left[ \kappa^3 r^6 \tilde{V}_0^2 + \kappa^3 r^6 \tilde{V}_0 (C + 1) - 2 C^2 \kappa^3 r^6 + b \tilde{Q}^2\tilde{V}_0^3 (\tilde{V}_0 - C) \right],
\end{equation}
where
\begin{equation}
	B=\frac{2 \sqrt{2} \pi^2}{\kappa^5 r^3 \tilde{V}_0^{\frac{5}{2}} \epsilon_{\tilde{V}} \gamma^{\frac{3}{2}} \sqrt{3(\tilde{V}_0 - C)}}>0,\quad b=30\kappa^3 - 18\kappa^2\lambda - 144\kappa\lambda^2 + 120\lambda^3.
\end{equation}
We can investigate the condition under which the action decreases with the parameter \(\lambda\), as a lower action exponentially enhances the probability. Therefore, we impose the condition \(\partial S_{E}^{\text{thick}}/\partial \lambda < 0\), where the requirement \(b < 0\) is crucial to ensure this inequality holds. The constraint \(b < 0\) yields the result $\lambda < \frac{7-\sqrt{249}}{20}\kappa \approx -0.439\kappa$.

To maintain the existence of an unstable maximum in the action for the thick-wall case ($a_{min} \approx a_{max}$), the second derivative in Eq.~\eqref{eq:47} must remain negative. This requirement leads to a more restrictive constraint, $\lambda < -\kappa$. Crucially, this condition does not contradict our previous findings; instead, it represents a systematic narrowing of the parameter space. It inherently satisfies not only the thin-wall requirement ($\lambda < 0$) and the foundational bound ($\lambda < \kappa/2$), but also the condition required to decrease the action in the thick-wall regime ($\lambda < \frac{7-\sqrt{249}}{20}\kappa \approx -0.439\kappa$). Therefore, $\lambda < -\kappa$ emerges as the most robust and unifying constraint for the theory.

To conclude our analysis in this section, by synthesizing the results from both the $a_{min} \ll a_{max}$ and $a_{min} \approx a_{max}$ scenarios, we establish a consistent parameter space for $\lambda$. The transition from the baseline requirement $\lambda < \kappa/2$ to the final condition $\lambda < -\kappa$ reflects the necessity of satisfying different physical criteria, specifically minimizing the action and ensuring potential instability, across all evolutionary regimes. This refined parameter range ensures that the $F(R,T)$ framework reliably supports sustained inflation, effectively addressing the duration problem inherent in the standard no-boundary proposal.

\section{Conclusion}\label{sec:5}
In this study, we investigate axion–dilaton wormhole solutions within the framework of $F(R,T)$ gravity, with the goal of addressing the issue of the insufficient number of  inflationary e-folds in the no-boundary proposal. Beginning with the GS-type and expanding wormhole solutions in an asymptotically flat Euclidean spacetime, our results demonstrate the distinct dynamical effects of the matter-geometry coupling. Based on the fundamental branch characterized by the minimum initial field $\phi_{0,min}$, we establish a negative correlation between the coupling parameter $\lambda$ and the initial scale factor $a_0$, as shown in Fig.~\ref{fig:diflammin}. This relationship indicates that a larger coupling parameter corresponds to a smaller initial wormhole throat. While the Euclidean action of this fundamental non-oscillatory branch increases with $\lambda$, we find that for the oscillatory branches, the coupling $\lambda$ significantly enhances the oscillatory behavior of the solutions. This amplified oscillation effectively lowers the overall Euclidean action compared to the standard $\lambda=0$ case, thereby enhancing the nucleation probability of these complex wormhole configurations.

Then, we can apply this framework to a ``wineglass'' half-wormhole model in Euclidean AdS spacetime. The wormhole evolution exhibits two characteristic scales, namely \( a_{\text{min}} \) and \( a_{\text{max}} \), as shown in Fig.~\ref{fig:a_EADS}. To compute the on-shell action analytically, we focus on two complementary limiting cases: \( a_{\text{min}} \ll a_{\text{max}} \) and \( a_{\text{min}} \approx a_{\text{max}} \). Our analysis of these two scenarios leads to the condition \( \lambda < -\kappa \), which  introduces an unstable maximum in \( V \) and concurrently reduces the value of \( S_E \). The presence of the unstable maximum alters the probability distribution of the initial states, making the evolution of universes from high-potential regions more probable. Within the no-boundary proposal, this model significantly enhances the probability of cosmological evolution paths that undergo prolonged inflation, offering a potential resolution to the short-duration inflation problem.

In summary, by utilizing the matter–geometry coupling within $F(R,T)$ gravity, our work presents a method that resolves the no-boundary proposal with the requirement for sustained inflation. The framework adjusts the probability distribution to favor initial conditions at high potential energies, allowing for a sufficient number of e-folds while maintaining theoretical consistency. This approach provides a valuable perspective on the  evolution of the early universe.

\section*{Acknowledgments}
The authors express their gratitude to George Lavrelashvili, Guangzhou Guo, Yizhi Liang, and Yigao Liu for their valuable suggestions and opinions, which have contributed significantly to the completion of this article. This work is supported by the National Natural Science Foundation of China (NSFC) with Grants No. 12175212. And it is finished on the server from Kun-Lun in College of Physics, Sichuan University.

\appendix
\section{Derivation of Constraints in the Thin-wall and Thick-wall Limits}\label{app1}

To establish a rigorous analytical framework for the wormhole throat geometry, we utilize the dimensionless parameters introduced in the main text: $\alpha \equiv 1+\frac{\lambda}{\kappa}$, $\delta \equiv 1-\frac{2\lambda}{\kappa}$, and $\gamma \equiv 1-\frac{\lambda}{\kappa}$, alongside the rescaled potential $\tilde{V}_0 = \frac{\kappa V_0}{3}$ and axion charge $\tilde{Q}^2 = \frac{\kappa Q^2}{3}$. In terms of these variables, the Friedmann constraint evaluated at the throat reduces to a cubic equation with respect to the squared scale factor $x \equiv a^2$:
\begin{equation}
	\delta\tilde{V}_0 x^3 - x^2 + \alpha\tilde{Q}^2 = 0.
\end{equation}

We denote the three real roots of this equation by $x_0$ ($a_{\max}^2$) and $x_1$ ($a_{\min}^2$), corresponding to the positive physical scales with $x_0 > x_1$, and $x_2$, representing the non-physical negative root. By virtue of Vieta's formulas, these roots are subject to the following algebraic relations:
\begin{align}
	x_0 + x_1 + x_2 &= \frac{1}{\delta\tilde{V}_0}, \label{eq:app_v1} \\
	x_0x_1 + x_1x_2 + x_2x_0 &= 0, \label{eq:app_v2} \\
	x_0x_1x_2 &= -\frac{\alpha\tilde{Q}^2}{\delta\tilde{V}_0}. \label{eq:app_v3}
\end{align}

To formulate constraints exclusively in terms of the physical characteristic scales, it is necessary to decouple the negative root $x_2$. From Eq.~\eqref{eq:app_v2}, we obtain $x_2 = -x_0x_1/(x_0 + x_1)$. Substituting this relation into Eq.~\eqref{eq:app_v1} and Eq.~\eqref{eq:app_v3} yields a reduced system of equations governing the dual positive scales:
\begin{align}
	\frac{x_0^2 + x_0x_1 + x_1^2}{x_0 + x_1} &= \frac{1}{\delta\tilde{V}_0}, \label{eq:app_exact1} \\
	\frac{x_0^2x_1^2}{x_0 + x_1} &= \frac{\alpha\tilde{Q}^2}{\delta\tilde{V}_0}. \label{eq:app_exact2}
\end{align}

The thin-wall limit is geometrically characterized by the condition $a_{\min}/a_{\max} \ll 1$, which translates to $x_1 \ll x_0$. In this asymptotic regime, Eq.~\eqref{eq:app_exact1} simplifies at leading order to $x_0 \approx 1/(\delta\tilde{V}_0)$. This limit corresponds to the parametric condition $\theta \to 0$ (or $\cos\theta \to 1$), indicating that $\alpha\tilde{Q}^2 \to 0$. Given the asymptotic behavior $a_{\max} \approx 1/\sqrt{\delta\tilde{V}_0}$, the defining thin-wall condition $a_{\min}/a_{\max} \ll 1$ is naturally equivalent to $a_{\min}\sqrt{\delta\tilde{V}_0} \ll 1$.

Conversely, the thick-wall limit is characterized by the degeneracy of the two physical scales, $a_{\min} \approx a_{\max}$, which implies $x_1 \approx x_0 \equiv x$. Under this degeneracy condition, Eq.~\eqref{eq:app_exact1} reduces to $3x/2 = 1/(\delta\tilde{V}_0)$, determining the degenerate root to be $x = 2/(3\delta\tilde{V}_0)$. Letting $\bar{a}$ denote the average scale factor between $a_{\min}$ and $a_{\max}$, we obtain $a_{\min} \approx a_{\max} \approx \bar{a} = \sqrt{x}$. This precisely reproduces the parameterization $\bar{a} = r/\sqrt{\delta\tilde{V}_0}$ utilized in the main text, fixing the numerical constant to be $r = \sqrt{2/3}$ and rigorously confirming that $r \sim \mathcal{O}(1)$.

\bibliographystyle{elsarticle-num}
\bibliography{ref} 
\end{document}